\documentclass[twoside,11pt]{article}

\usepackage{jmlr2e}

\usepackage{hyperref}
\usepackage{url}
\usepackage{subcaption}
\usepackage[utf8]{inputenc} 
\usepackage[T1]{fontenc}    
\usepackage{url}            
\usepackage{xcolor} 
\usepackage{amsfonts}       
\usepackage{nicefrac}       
\usepackage{float}
\usepackage{wrapfig}
\usepackage[ruled,vlined,linesnumbered]{algorithm2e}
\usepackage{algpseudocode}
\usepackage{color, colortbl}
\usepackage{enumitem}
\usepackage{caption}
\usepackage{todonotes}
\usepackage{mdframed}
\usepackage[skins]{tcolorbox}
\definecolor{Gray}{gray}{0.9}
\definecolor{LightCyan}{rgb}{0.88,1,1}
\usepackage{subcaption}
\usepackage{graphicx}      
\usepackage{natbib}        

\usepackage{microtype}
\usepackage{booktabs} 

\usepackage{hyperref}

\usepackage{mathtools}
\usepackage{amsthm}

\newtheorem{example}{Example} 
\newtheorem{theorem}{Theorem}
\newtheorem{lemma}[theorem]{Lemma} 
 
\newtheorem{remark}[theorem]{Remark}

\newtheorem{definition}[theorem]{Definition}

\definecolor{colorone}{rgb}{0.00,0.45,0.74}
\definecolor{colortwo}{rgb}{0.85,0.33,0.10}
\definecolor{colorthree}{rgb}{0.49,0.18,0.56}
\definecolor{colorfour}{rgb}{0.93,0.69,0.13}
\definecolor{colorfive}{rgb}{0.47,0.67,0.19}
\definecolor{colorsix}{rgb}{0.30,0.75,0.93}

\usepackage{lastpage}

\firstpageno{1}

{}

\theoremstyle{definition}

\newcommand{\statee}{\mathbf{s}}
\newcommand{\timeid}{k}
\newcommand{\states}{\mathcal{S}}
\newcommand{\controls}{\mathcal{C}}
\newcommand{\reals}{\mathbb{R}}

\newcommand{\integers}{\mathbb{Z}}
\newcommand{\posint}{\integers^{\ge 0}}

\newcommand{\dynamics}{\mathbf{f}}

\newcommand{\action}{\mathbf{a}}

\newcommand{\policy}{\pi}
\newcommand{\param}{\mathbf{\theta}}
\newcommand{\params}{\Theta}
\newcommand{\optparam}{\param^\star}
\newcommand{\optpolicy}{\pi_{\optparam}}
\newcommand{\mypara}[1]{\vspace{0.3em} \noindent {\bf #1}.}
\newcommand{\myipara}[1]{\vspace{0.4em} \noindent {\bf #1}.}
\newcommand{\traj}{\sigma}
\newcommand{\init}{\mathcal{I}}
\newcommand{\sampledinits}{\widehat{\init}}

\newcommand{\horizon}{K}

\newcommand{\alw}{\mathbf{G}}
\newcommand{\ev}{\mathbf{F}}
\newcommand{\unt}{\mathbf{U}}
\newcommand{\intvl}{I}

\newcommand{\relu}{\mathsf{ReLU}}

\newcommand{\stltonn}{\mathsf{STL2NN}}
\newcommand{\lbfortl}{\mathsf{LB4TL}}

\newcommand{\gradient}[2]{\nabla_{#2} #1}

\newcommand{\softplus}{\mathsf{softplus}}
\newcommand{\swish}{\mathsf{swish}}
\newcommand{\stltolb}{\tilde{\rho}}

\newcommand{\ignore}[1]{}

\newcommand{\muc}[2]{\multicolumn{#1}{c}{#2}}

\newcommand{\until}{\mathbf{U}}

\newcommand{\adam}{\mathsf{Adam}}

\newcommand{\samplednospot}{\mathsf{smpl}\!\left(\traj[\statee_0\ ;\param], \mathcal{T}\right)}
\newcommand{\samplednospotj}{\mathsf{smpl}\!\left(\traj[\statee_0\ ;\param^{(j)}], \mathcal{T}\right)}
\newcommand{\samplednospotf}{\mathsf{smpl}\!\left(\traj[\statee_0\ ;\param_{1}], \mathcal{T}^1\right)}
\newcommand{\samplednospotfwp}{\mathsf{smpl}\!\left(\traj[\statee_0\ ;\param_{2}], \mathcal{T}^2\right)}

\newcommand{\sampledspotq}{\mathsf{smpl}\!\left(\traj[\statee_0\ ;\param], \mathcal{T}^q\right)}
\newcommand{\sampledspotqj}{\mathsf{smpl}\!\left(\traj[\statee_0\ ;\param^{(j)}], \mathcal{T}^q\right)}
\newcommand{\sampledspotqs}{\mathsf{smpl}\!\left(\traj[\statee_0\ ;\param_{3}], \mathcal{T}^q\right)}

\newcommand{\subnospotj}{\mathsf{sub}\!\left(\trajsinitj, \mathcal{T}\right)}
\newcommand{\subspotqj}{\mathsf{sub}\!\left(\trajsinitj, \mathcal{T}^q\right)}

\newcommand{\trajsinit}{\traj[\statee_0\ ;\param]}
\newcommand{\trajsinitj}{\traj[\statee_0\ ;\param^{(j)}]}

\newcommand{\rob}{\rho}

\newcommand{\release}{\mathbf{R}}

\newcommand{\navid}[1]{\textcolor{black}{#1}}
\newcommand{\navidd}[1]{\textcolor{black}{#1}}

\newcommand{\navidddd}[1]{\textcolor{black}{#1}}
\newcommand{\navidg}[1]{\textcolor{black}{#1}}

\begin{document}

\title{Scaling Learning based Policy Optimization for Temporal Logic Tasks by Controller Network Dropout}
\author{\name Navid Hashemi     \email navidhas@usc.edu \\
       \addr University of Southern California, Los Angeles, California, United States\\
       \AND
       \name Bardh Hoxha        \email Bardh.Hoxha@toyota.com \\
       \addr Toyota NA R\&D, Ann Arbor, Michigan, Unites States\\
       \AND
       \name Danil Prokhorov    \email Danil.Prokhorov@toyota.com \\
       \addr Toyota NA R\&D, Ann Arbor, Michigan, Unites States\\
       \AND
       \name Georgios Fainekos  \email Georgios.Fainekos@toyota.com \\
       \addr Toyota NA R\&D, Ann Arbor, Michigan, Unites States\\
       \AND
       \name Jyotirmoy Deshmukh \email jdeshmuk@usc.edu\\
       \addr University of Southern California, Los Angeles, California, United States\\
       }

\maketitle
\begin{abstract}                
This paper introduces a model-based approach for training feedback controllers for an autonomous agent operating in a highly nonlinear (albeit deterministic) environment. We desire the trained policy to ensure that the agent satisfies specific task objectives and safety constraints, both expressed in Discrete-Time Signal Temporal Logic (DT-STL). One advantage for reformulation of a task via formal frameworks, like DT-STL, is that it permits quantitative satisfaction semantics. In other words, given a trajectory and a DT-STL formula, we can compute the {\em robustness}, which can be interpreted as an approximate signed distance between the trajectory and the set of trajectories satisfying the formula. We utilize feedback control, and we assume a feed forward neural network for learning the feedback controller. We show how this learning problem is similar to training recurrent neural networks (RNNs), where the number of recurrent units is proportional to the temporal horizon of the agent's task objectives. This poses a challenge: RNNs are susceptible to vanishing and exploding gradients, and na\"{i}ve gradient descent-based strategies to solve long-horizon task objectives thus suffer from the same problems. To tackle this challenge, we introduce a novel gradient approximation algorithm based on the idea of dropout or gradient sampling. One of the main contributions is the notion of {\em controller network dropout}, where we approximate the NN controller in several time-steps in the task horizon by the control input  obtained using the controller in a previous training step. We show that our control synthesis methodology, can be quite helpful for stochastic gradient descent to converge with less numerical issues, enabling scalable backpropagation over long time horizons and trajectories over high dimensional state spaces. We demonstrate the efficacy of our approach on various motion planning applications requiring complex spatio-temporal and sequential tasks ranging over thousands of time-steps.
\end{abstract}


\section{Introduction}
\label{sec:intro}

The use of \navid{Neural Networks (NN)} for feedback control enables data-driven control design for highly nonlinear environments. 
\navidg{The literature about training NN-based controllers or neuro-controllers is plentiful, e.g., see \cite{berducci2021hierarchical,li2017reinforcement, chua2018deep,fang2019survey, hashimoto2022stl2vec, Lewis2007}}. 
Techniques to synthesize neural controllers (including deep
RL methods) largely focus on optimizing cost functions that are constructed from user-defined state-based rewards or costs.
These rewards are often proxies for desirable long-range behavior of
the system and can be error-prone \cite{pan2022effects,skalse2022defining,amodei2016concrete} and
often require careful design \cite{hadfield2017inverse,sorg2010reward}.

On the other hand, in most engineered safety-critical systems, the
desired behavior can be described by a set of spatio-temporal
task-objectives, \navidg{e.g., \cite{fainekos2009temporal,belta2017formal}}.  For example, consider modeling a mobile robot where the system
must reach region $R_1$ before reaching region $R_2$, while avoiding
an obstacle region.  Such spatio-temporal task objectives can be
expressed in the mathematically precise and symbolic formalism of
\navidg{Discrete-Time variant (DT-STL) \cite{FainekosP06fates} of Signal Temporal Logic (STL) \cite{maler2004monitoring} }. A key advantage of DT-STL is that for any
DT-STL specification and a system trajectory, we can efficiently
compute the {\em robustness degree}, i.e., the approximate signed
distance of the trajectory from the set of trajectories
\navidg{satisfying/violating the specification
\citep{donze2010robust,FainekosP06fates}}.

Control design with DT-STL specifications using the robustness degree as an
objective function to be optimized is \navidg{an approach} that brings together
two separate threads: (1) smooth approximations to the robustness degree of STL
specifications \citep{gilpin2020smooth,pant2017smooth} enabling the use of STL
robustness in gradient-based learning of open-loop control policies, and (2)
representation of the robustness as a computation graph allowing its use in
training neural controllers using back-propagation
\cite{YaghoubiF2019tecs,leung2019backpropagation, hashemi2023neurosymbolic, HashemiEtAl2023acc}.
While existing methods have demonstrated some success in training open-loop NN
policies \cite{leung2019backpropagation,LeungAP2021wafr}, and also closed-loop
NN policies
\cite{hashemi2023neurosymbolic, HashemiEtAl2023acc, YaghoubiF2019tecs}, several
key limitations still remain. Consider the problem of planning the trajectory of
a UAV in a complex, GPS-denied urban environment; here, it is essential that the
planned trajectory span several minutes while avoiding obstacles and \navidg{reaching
several sequential goals \citep{windhorst2021strategic,pant2021co, vachtsevanos2005mission}}. However, none of the
existing methods to synthesize closed-loop (or even open-loop) policies scale to
handle long-horizon tasks. 

A key reason for this is the inherent computational challenge in dealing with
long-horizon specifications. Training open-loop policies treats the sequence of
optimal control actions over the trajectory horizon as decision variables to
maximize the robustness of the given STL property. Typical approaches use
gradient-descent where in each iteration, the new control actions (i.e. the
open-loop policy) are computed using the gradient of the DT-STL property w.r.t.
the control actions. If the temporal horizon of the DT-STL property is
$\horizon$, then, this in turn is computed using back-propagation of the DT-STL
robustness value through a computation graph representing the composition of the
DT-STL robustness computation graph and $\horizon$ copies of the system
dynamics. Similarly, if we seek to train closed-loop (neural) feedback control
policies using gradient descent, then we can treat the one-step environment
dynamics and the neural controller as a recurrent unit that is repeated as many
times as the temporal horizon of the DT-STL property. Gradient updates to the
neural controller parameters are then done by computing the gradient of the STL
computation graph composed with this RNN-like structure. In both cases, if the
temporal horizon of $\varphi$ is several hundred steps, then gradient
computation requires back-propagation through those many steps. These procedures
are quite similar to the ones used for training an RNN with many recurrent
units. It is well-known that back-propagation through RNNs with many recurrent
units faces problems of vanishing and exploding gradients
\citep{goodfellow2016deep,ba2016layer}. To address these limitations, we propose
a sampling-based
approximation of the gradient of the objective function (i.e. the STL
property), that is particularly effective when dealing with behaviors
over large time-horizons. Our key idea is to approximate the gradient during back-propagation by an approximation scheme similar to the idea of \navid{dropout} layers used \navidg{in deep neural networks \cite{srivastava2014dropout}}. The main idea of \navid{dropout} layers is to probabilistically set the output of some neurons in the layer to zero in order to prevent over-fitting. \navidg{We do a similar trick: in each training iteration we pick some recurrent units to be "frozen", i.e., we use older fixed values of the NN parameters for the frozen layers, effectively approximating the gradient propagation through those layers.} We show that this can improve training of NN
controllers by at least an {\em order of magnitude}.  Specifically, we
reduce training times from hours to minutes, and can also train
reactive planners for task objectives that have large time horizons.

\noindent To summarize, we make the following contributions:
\begin{enumerate}[nosep,leftmargin=1.5em]
\item We develop a sampling-based approach, inspired by dropout \cite{srivastava2014dropout}, to approximate the gradient of DT-STL robustness w.r.t. the 
NN controller parameters. Emphasizing the time-steps that contribute the most to the gradient, our method randomly samples time points over the trajectory. We utilize the 
structure of the STL formula and the current system trajectory to decide which time-points 
represent critical information for the gradient.
\item \navidg{We develop a back-propagation method that uses a combination of the proposed sampling approach and the smooth version of the robustness degree of a DT-STL specification to train NN controllers.}
\item  We demonstrate the efficacy of our approach on high dimensional nonlinear dynamical
systems involving long-horizon and complex temporal specifications.
\end{enumerate}






%
%
%
\
\subsection{Organization and Notations}

The rest of the paper is organized as follows. In
Section~\ref{sec:prelim}, we introduce the notation and the problem
definition. We propose our learning-based control synthesis algorithms
in Section~\ref{sec:training}, present experimental evaluation in
Section~\ref{sec:experiments}, and conclude in Section~\ref{sec:conc}. We use bold letters to indicate vectors and vector-valued functions, and calligraphic letters to denote sets. A feed forward neural network (NN) with $\ell$ hidden layers is denoted by the vector $[n_0,n_1,\cdots n_{\ell+1}]$, where $n_0$ denotes the number of inputs, $n_{\ell+1}$ is the number of outputs and for all $i \in 1,2,\cdots,\ell$, and $n_i$ denotes the width of $i^{th}$ hidden layer. The notation $x \stackrel{u}{\sim} \mathcal{X}$ implies the random variable $x$ is sampled uniformly from the set $\mathcal{X}$.

\section{Preliminaries}
\label{sec:prelim}
\mypara{NN Feedback Control Systems (NNFCS)}
Let $\statee$ and $\action$ denote the state and action variables that take values from compact sets $\states \subseteq \reals^n$ and $\controls \subseteq \reals^m$, respectively. We use $\statee_\timeid$ (resp. $\action_\timeid$) to denote the value of the state (resp. action) at time \navid{$\timeid\in
\posint$}. We define a neural network controlled system (NNFCS) as a  recurrent difference equation
\begin{equation}
\label{eq:nncs}
    \statee_{\timeid+1} = \dynamics(\statee_\timeid,\action_\timeid), 
\end{equation}
where $\action_\timeid = \policy_\param(\statee_\timeid, \timeid)$ is the control policy.
We assume that 
the control policy is a parameterized function $\policy_\param$, where $\param$ is a vector of parameters that takes values in $\params$. Later in the paper, we instantiate the specific parametric form using a neural network for the controller. That is, given a fixed vector of parameters $\param$, the parametric control policy $\policy_\param$ returns an action $\action_\timeid$ as a function of the current state $\statee_\timeid \in \states$ and time \navid{$\timeid$}, i.e., $\action_\timeid = \policy_\param(\statee_\timeid, \timeid)$\footnote{\navid{Our proposed feedback policy explicitly uses time as an input. This approach is motivated by the need to satisfy temporal tasks, which requires time awareness for better decision-making.}}.

\begin{figure}[t]
\centering
    \includegraphics[width=0.6\linewidth]{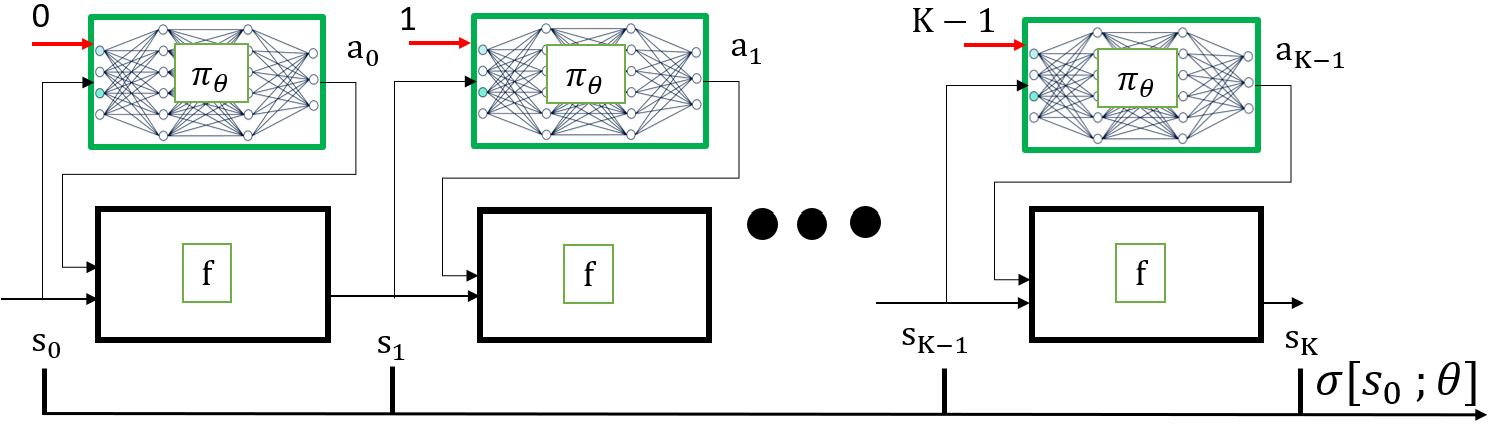}
    \caption{Shows an illustration of the recurrent structure for the control feedback system.}
    \label{fig:recurrent}
\end{figure}

\mypara{Closed-loop Model Trajectory} For a discrete-time \navid{NNFCS} as shown in Eq.~\eqref{eq:nncs}, and a set of designated initial states $\init \subseteq \states$, under a pre-defined feedback policy $\policy_\param$, Eq.~\eqref{eq:nncs} represents an autonomous discrete-time dynamical system. For a given initial state $\statee_0 \in \init$, a system trajectory $\trajsinit$ is a function mapping
time instants \navid{$\timeid \in 0,1,\cdots,\horizon$ to $\states$, where $\trajsinit(\timeid) = \statee_\timeid$, and for all $\timeid \in 0,1,\cdots,\horizon-1$, $\statee_{\timeid+1} = \dynamics(\statee_\timeid,\pi_\param(\statee_\timeid, \timeid))$. In case the dependence to $\param$ is obvious from the context, we utilize the notation $\statee_\timeid$ to refer to $\trajsinit(\timeid)$.}
Here, $\horizon$ is some integer called the trajectory horizon, and the exact value of $\horizon$ depends on the DT-STL task objective that the closed-loop model trajectories must satisfy. The computation graph for this trajectory is a recurrent structure. Figure \ref{fig:recurrent} shows an illustration of this structure and its similarity to an RNN. 

\mypara{Task Objectives and Safety Constraints} We assume that task objectives and safety constraints are specified using the syntax of \navidg{Discrete-Time variant (DT-STL) \cite{FainekosP06fates} of Signal Temporal Logic (STL) \cite{maler2004monitoring} }. We assume that DT-STL formulas are specified in positive normal form, i.e., all negations are pushed to the signal predicates \footnote{Any formula in DT-STL can be converted to a formula in positive normal form using DeMorgan's laws and the duality between the Until and Release operators)}
\begin{equation}
\label{eq:stlfrag}
\varphi\  =\  h(\statee) \bowtie 0 \mid 
             \varphi_1 \wedge \varphi_2 \mid
             \varphi_1 \vee \varphi_2 \mid \\
             \varphi_1 \until_\intvl \varphi_2 \mid \varphi_1 \release_\intvl \varphi_2 
\end{equation}
where $\until_\intvl$ and $\release_\intvl$ are the timed until and release operators, $\bowtie \in \{ \le, <, >, \ge \}$, and $h$ is a function from $\states$ to $\reals$.
In this work, since we use discrete-time semantics for STL (referred to as DT-STL), the time interval $\intvl$ is a bounded interval of integers, i.e., $\intvl = [a,b], a \leq b $. The timed eventually ($\ev_\intvl$) and always ($\alw_\intvl$) operators can be syntactically defined through until and release. That is, $\ev_\intvl\varphi \equiv \top \until_\intvl \varphi$ and $\alw_\intvl\varphi \equiv \bot \release_\intvl \varphi$ where $\top$ and $\bot$ represent true and false. The formal semantics of DT-STL over discrete-time trajectories have been previously presented in \citep{FainekosP06fates}. We briefly recall them here.

\mypara{Boolean Semantics and Formula Horizon} We denote the formula $\varphi$ being true at time $\timeid$ in trajectory $\trajsinit$ by $\trajsinit,\timeid \models \varphi$. We say that $\trajsinit,\timeid \models h(\statee) \bowtie 0$ iff $h(\trajsinit(\timeid)) \bowtie 0$. The semantics of the Boolean operations ($\wedge$, $\vee$) follow standard logical semantics of conjunctions and disjunctions, respectively. For temporal operators, we say $\trajsinit, \timeid \models \varphi_1 \until_\intvl \varphi_2$ is true if there is a time $\timeid'$, s.t. $\timeid' - \timeid \in \intvl$ where $\varphi_2$ is true and for all times $\timeid'' \in [\timeid,\timeid')$, $\varphi_1$ is true. Similarly, $\trajsinit, \timeid \models \varphi_1 \release_\intvl \varphi_2$ is true if for all times $\timeid'$ with $\timeid'-\timeid \in \intvl$, $\varphi_2$ is true, or there exists some time $\timeid'' \in [\timeid,\timeid')$ such that $\varphi_1$ was true. The temporal scope or horizon of a DT-STL formula defines the \navidddd{last time-step} required to evaluate the formula, $\trajsinit, 0 \models \varphi$ (see \citep{maler2004monitoring}). For example, the temporal scope of the formula $\ev_{[0,3]}(x>0)$ is $3$, and that of the formula $\ev_{[0,3]}\alw_{[0,9]}(x>0)$ is $3 + 9 = 12$. We also set the horizon of trajectory equivalent to the horizon of formula, as we plan to monitor the satisfaction of the formula by the trajectory.

\mypara{Quantitative Semantics (Robustness value) of DT-STL}  Quantitative semantics of DT-STL roughly define a signed distance of a given trajectory from the set of trajectories satisfying or violating the given DT-STL formula. There are many alternative semantics proposed in the literature \citep{donze2010robust,FainekosP06fates,rodionova2022combined,akazaki2015time}; \navidg{in this paper, we focus on the semantics from \citep{donze2010robust} that are shown in Table \ref{eq:sem}}. The robustness value $\rob(\varphi,\trajsinit,\timeid)$ of a DT-STL formula $\varphi$ over a trajectory $\trajsinit$ at time $\timeid$ is defined \navid{recursively as reported in Table \ref{eq:sem}}\footnote{For brevity, we omit the trajectory from the notation, as it is obvious from the context.}.
\begin{table}[t]
\centering
\begin{equation*}
\resizebox{\linewidth}{!}{$
\begin{array}{cc|cc}
\toprule
\varphi & \rob(\varphi,\timeid) & \varphi & \rob(\varphi,\timeid)\\
\hline
h(\statee_\timeid) \ge 0 &
h(\statee_\timeid) & \ev_{[a,b]}\psi &
\displaystyle \max_{\timeid' \in  [\timeid+a,\timeid+b]} \rob(\psi,\timeid')\\
\varphi_1 \wedge \varphi_2 &
\displaystyle \min(\rob(\varphi_1,\timeid),\rob(\varphi_2,\timeid)) &  \varphi_1 \unt_{[a,b]} \varphi_2 &
\displaystyle \max_{\substack{\timeid' \in [\timeid+a,\timeid+b]}}\left( \min\left(\rob(\varphi_2,\timeid'),\displaystyle\min_{\substack{\timeid'' \in  [\timeid,\timeid')}} \rob(\varphi_1,\timeid'')\right) \right)  \\
\varphi_1 \vee \varphi_2 &
\displaystyle \max(\rob(\varphi_1,\timeid),\rob(\varphi_2,\timeid)) &  \varphi_1 \release_{[a,b]} \varphi_2 &
\displaystyle \min_{\timeid' \in [\timeid+a,\timeid+b]} \left(
\max\left( \rob(\varphi_2,\timeid'),\displaystyle\max_{\timeid'' \in [\timeid,\timeid')} \rob(\varphi_1,\timeid'')\right) \right)  \\
\alw_{[a,b]}\psi &
\displaystyle \min_{\timeid' \in [\timeid+a,\timeid+b]} \rob(\psi,\timeid') &  &  \\
\bottomrule
\end{array}
$}
\end{equation*}
\caption{Quantitative Semantics of STL\label{eq:sem}}
\end{table}
We note that if $\rob(\varphi,\timeid) > 0$ the DT-STL formula $\varphi$ is satisfied at time $\timeid$, and we say that the formula $\varphi$ is satisfied by a trajectory if $\rob(\varphi,0) > 0$.




\mypara{Prior Smooth Quantitative Semantics for DT-STL} To address non-differentiability of the robust semantics of STL, there have been a few alternate definitions of smooth approximations of the robustness in the literature. The initial proposal for this improvement is provided by \cite{pant2017smooth}. Later the authors in \cite{gilpin2020smooth} proposed another smooth semantics which \navidg{in addition is a guaranteed} lower bound for the robustness value that can be even more advantageous computationally. We denote the smooth robustness of trajectory $\traj[\statee_0\ ;\param]$ for temporal specification $\varphi$,  with $\stltolb(\varphi, \traj[\statee_0\ ;\param], 0)$.

\mypara{ Neuro-Symbolic Smooth Semantics} The prior smooth semantics for gradient computation over DT-STL \citep{gilpin2020smooth,pant2017smooth,leung2019backpropagation} perform backward computation on a computation graph that is generated based on dynamic programming. Although these computation graphs are efficient for forward computation, they may face computational difficulty for backward computation over the robustness when the specification is highly complex or its task horizon is noticeably long. Unlike the previous computation graphs that are based on dynamic programming, the neurosymbolic computation graph $\stltonn$ \cite{hashemi2023neurosymbolic}, directly utilizes the STL tree \cite{donze2010robust} to generate a feedforward $\relu$ neural network, whose depth grows logarithmically with the complexity of specification. This makes back-propagation more feasible for complex specifications. On the other hand, the way it formulates the robustness (Feedforward NN) facilitates the back-propagation process, by enabling vectorized gradient computation. However, since $\stltonn$ is exactly identical to the non-smooth robustness introduced in Table~\ref{eq:sem}, \navidg{we proposed in \cite{hashemi2024scaling} a smooth} under-approximation for $\stltonn$ replacing the $\relu$ activation function with $\swish()$ and $\softplus()$, and introduced this as $\lbfortl$, and here we utilize this smooth semantics.

\mypara{Problem Definition} In this paper, we provide \navid{model-based} algorithms to learn a  policy $\optpolicy$ that maximizes the degree to which certain task objectives and safety constraints are satisfied. In particular, we wish to learn a neural network (NN) control policy $\policy_{\param}$ (or equivalently the parameter values $\param$), s.t. for any initial state $\statee_0 \in \init$, using the control policy $\policy_ \param$, the trajectory obtained, i.e., $\trajsinit$ satisfies a given DT-STL formula $\varphi$. In other words, our ultimate goal is to solve the optimization problem shown in
Eq.~\eqref{eq:opt1}.
For brevity, \navidg{ we use $F(\statee_\timeid,\timeid\ ;\param)$ to denote
$\dynamics\left(\statee_\timeid , \policy_{\param}\left(\statee_\timeid , \timeid \right) \right)$}.
\begin{equation}\label{eq:opt1}
\begin{aligned}
&\param^*  =  \arg\max_{\param} \left(  \navidg{ \min_{\statee_0 \in \init} \left[\rob(\varphi, \trajsinit, 0) \right] } \right),\\
& \textbf{s.t.}\ \forall(\timeid\in \mathbb{Z} \wedge 1\le \timeid < K):  \statee_{\timeid+1} = F(\statee_\timeid,\timeid\ ;\param). 
\end{aligned}
\end{equation}

\navidg{However, ensuring that the robustness value is positive for all \( \statee_0 \in \init \) is computationally challenging. Therefore, we relax the problem to maximizing the $\min$ value of the robustness only over a set of states $\hat \init$ sampled from the initial states $\init$, i.e., $\param^* \approx \arg\max_{\param} \left( \min_{\statee_0 \in \sampledinits} \left[\rob(\varphi, \trajsinit, 0) \right] \right)$. 
We solve this problem using algorithms based on stochastic gradient descent followed by statistical verification to obtain high-confidence control policies.}


\section{Training Neural Network Control Policies}
\label{sec:training}
\begin{figure}[t]
\centering
    \includegraphics[width=0.5\linewidth]{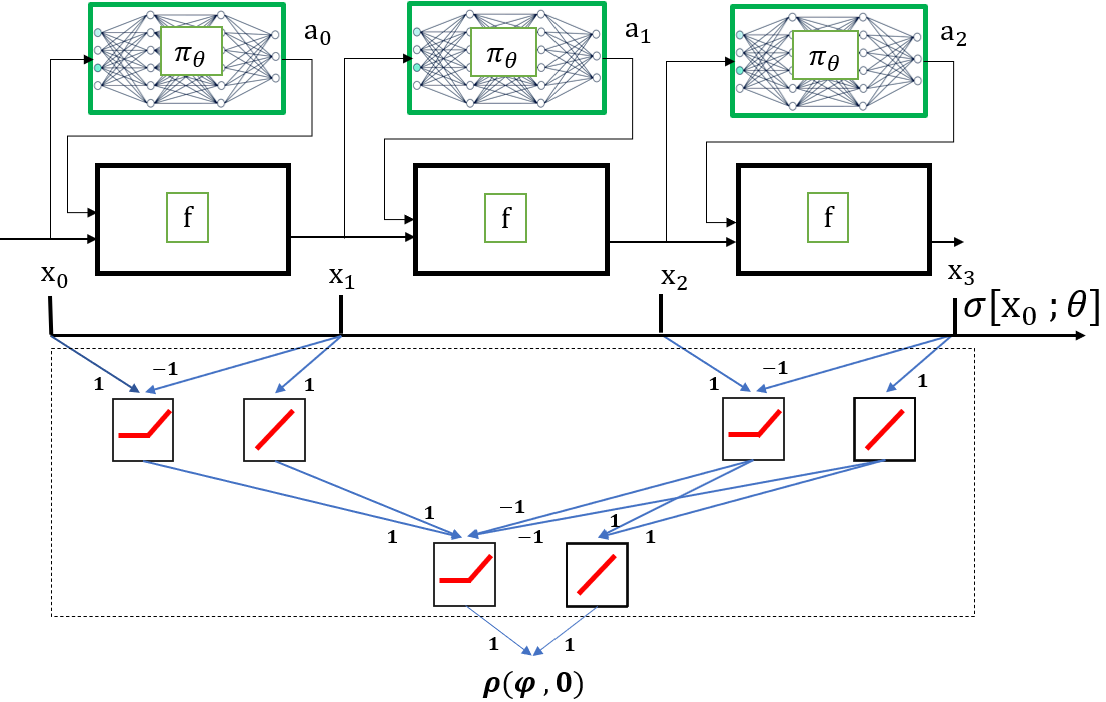}
    \caption{This figure shows the symbolic trajectory generated by NN feedback controller, and the computation graph for DT-STL robustness. The DT-STL robustness is presented as a Nero-symbolic computation graph \cite{hashemi2023neurosymbolic} via $\relu$ and Linear activation functions.}\vspace{-2mm}
    \label{fig:recurrentexample}
\end{figure}

Our solution strategy is to treat each time-step of the given dynamical equation in Eq.~\eqref{eq:nncs} as a recurrent unit. We then sequentially compose or unroll as many units as required by the horizon of the DT-STL specification. 
\begin{example}\label{ex:1} Assume a one-step dynamics with scalar state, $\mathbf{x} \in \mathbb{R} $ and scalar feedback control policy $\text{a}_\timeid = \policy_\param(\mathbf{x}_\timeid)$ as, $\mathbf{x}_{\timeid+1}=\dynamics(\mathbf{x}_\timeid,\policy_\param(\mathbf{x}_\timeid))$. If the specification is $\ev_{[0,3]}(\mathbf{x}>0)$, then, we use 3 instances of $\dynamics(\mathbf{x}_\timeid,\policy_\param(\mathbf{x}_\timeid))$ by setting the output of the $\timeid^{\text{th}}$ unit to be the input of the $(\timeid+1)^{\text{th}}$ unit. This unrolled structure implicitly contains the system trajectory, $\traj[\mathbf{x}_0,\ ;\param]$ starting from some initial state $\mathbf{x}_0$ of the system. The unrolled structure essentially represents the symbolic trajectory, where each recurrent unit shares the NN parameters of the controller (see Figure. \ref{fig:recurrentexample} for more detail). By composing this structure with the robustness semantics representing the given DT-STL specification $\varphi$; we have a computation graph that maps the initial state of the system in Eq.~\eqref{eq:nncs} to the robustness degree of $\varphi$.  Thus, training the parameters of this resulting structure to guarantee that its output is positive (for all initial states) guarantees that each system trajectory satisfies $\varphi$. 
\end{example}

However, we face a challenge in training the neural network controller that is embodied in this structure.

\noindent {\em Challenge:} \navidg{Since our computation graph resembles a recurrent structure with repeated units proportional to the formula's horizon, naïve gradient-based training algorithms struggle with gradient computation when using back-propagation through the unrolled system dynamics.}
In other word, the gradient computation faces the same issues of vanishing and
exploding gradients when dealing with long trajectories.

\mypara{Controller Synthesis as an Optimization problem}
In order to train the controller, we solve the following problem:
\begin{equation}\label{eq:trainopt}
\param^* =  {\arg  \max}_{\param} \left(   \min_{\statee_0 \in \sampledinits}\left[\rob(\varphi, \traj[\statee_0\ ;\param], 0) \right] \right), \qquad
\textbf{s.t.}\ \  \traj[\statee_0\ ;\param](\timeid+1) = F(\statee_\timeid,\timeid\ ;\param).
\end{equation}
We thus wish to maximize the expected value of the robustness for trajectories
starting in states uniformly sampled from the set of initial states. An
approximate solution for this optimization problem is to train the NN controller
using a vanilla back-propagation algorithm to compute the gradient of the
objective function for a subset of randomly sampled initial states
$\sampledinits \subset \init$, and updates the parameters of the neural network
controller using this gradient.

\begin{remark} A training-based solution to the optimization problem does not
guarantee that the specification is satisfied {\em for all} initial states
$\statee_0 \in \init$. \navidg{To tackle this, we can use a methodology like
\cite{hashemi2023neurosymbolic} that uses reachability analysis to verify the
synthesized controller. However, given the long time-horizon, this method may
face computational challenges.} An \navidg{alternative} approach is to eschew deterministic
guarantees, and instead obtain probabilistic guarantees (see Sec.
\ref{sec:verific}). \end{remark}
\section{Extension to Long Horizon Temporal Tasks \& Higher Dimensional Systems}
\label{sec:extension}

In this section, we introduce an approach to alleviate the problem of
exploding/vanishing gradients outlined in the previous section. Our solution
approach is inspired by the idea of using dropout layers
\citep{srivastava2014dropout} in training deep neural networks. In our approach,
we propose a sampling-based technique, where we only select certain time-points
in the trajectory for gradient computation, while using a fixed older control
policy at the non-selected points. Our approach to gradient sampling can be also
viewed through the lens of stochastic depth, as suggested by
\cite{huang2016deep}, which involves sampling layers followed by identity
transformations provided in ResNet. However, our methodology differs as we
employ a distinct approach that is better suited for control synthesis within
the Signal Temporal Logic (STL) framework. \navid{ Before starting our main discussion on this topic, we \navidg{first provide} an overview of this section,
\begin{itemize}[nosep,leftmargin=1.5em]
\item In section \ref{subsec:Sampledef}, we introduce the notion of gradient approximation through sampling the trace, and justify why it is a suitable replacement for the original gradient, in case the original gradient is not accessible (e.g. long-horizon tasks).
\item In section \ref{subsec:critic}, we put forward the notion of critical time which states that the robustness of DT-STL is only related to a specific time-step. We then propose the idea of including this time-step into  our gradient approximation technique.
\item In section \ref{subsec:saferesmoothing}, we bring up the point that gradient approximation using the critical time, may in some cases, result in failure for training. In these cases, we suggest approximating the DT-STL robustness as a function of all the trace, that is the smooth version of the robustness semantics.
\item In section \ref{subsec:approxcompute}, we explain how to approximate the gradient for both of the scenarios we proposed above (e.g., critical time \& smooth semantics). We also introduce Algorithm~\ref{algo:training} \navidg{which concludes} section \ref{sec:extension}.
\end{itemize}
}

\subsection{Sampling-Based Gradient Approximation Technique} \label{subsec:Sampledef}

We propose to sample random time-steps in the recurrent structure shown in
Fig.~\ref{fig:recurrent} and at the selected time-step, we do an operation that
is similar to dropping the entire neural controller. However, approximating the
gradient by dropping out the controller at several time-steps may result in
inaccurate approximation. We compensate for this by repeating our modified
dropout process and computing cumulative gradients. Restriction of dropout to
sample time-steps results in less number of self multiplication of weights and
therefore alleviates the problem of vanishing/exploding gradient. To ensure that
the trajectory is well-defined, when we drop out the controller unit at a
selected time-step, we replace it with a constant function that returns the
evaluation of the controller unit (at that specific time-step) in the forward
pass. We formalize this using the notion of a sampled trajectory in
Definition.~\ref{def:straj}.

\begin{definition}[Sub-trajectory \& Sampled trajectory]\label{def:straj}
    Consider the set of $N$ different sampled time-steps $\mathcal{T} = \left\{t_0=0, t_1, t_2, \cdots, t_N \right\}$ sampled from the horizon $\mathcal{K} = \left\{0, 1, 2, \cdots,\horizon \right\}$, and also the initial state $\statee_0$, and the control parameters ${\param^{(j)}}$ in the gradient step $j$. 
    The sub-trajectory, $\subnospotj = \statee_0, \statee_{t_1}, \statee_{t_2}, \cdots, \statee_{t_N} $ is simply a selection of $N$ states from $\traj[\statee_0\ ;\param^{(j)}]$ with time-steps $t_i \in \mathcal{T}$. 
    In other words, for all $i \in \left\{0,1, \cdots, N \right\}$: $\subnospotj(i) = \traj[\statee_0\ ;\param^{(j)}](t_i)$.
    Now, consider the sub-trajectory $\subnospotj$, and a sequence of actions $\action_0, \action_1, \cdots,\action_{\horizon-1}$ resulting from $\statee_0$ and $\param^{(j)}$. 
    For any $t_i \in \mathcal{T}$, we drop out the NN controllers on time steps $t_i+1, t_i+2, \cdots,t_{i+1}-1$ and replace them with the actions $\action_{1+t_i}, \action_{2+t_i}, \cdots \action_{t_{i+1}-1}$. 
    This provides a variant of sub-trajectory called {\it sampled trajectory}, and we denote it by $\samplednospotj$. In other words, for any time-step $t_i \in \mathcal{T}$, assuming the function
    $\dynamics_{i+1}: \mathcal{S}\times \Theta \to \mathcal{S}$ 
    (for brevity, henceforth, we denote $\dynamics_{i+1}(\statee\ ;\param^{(j)})$ by $\dynamics_{i+1}^{(j)}(\statee)$):
    $$
    \dynamics_{i+1}^{(j)}(\statee) = \dynamics(\dynamics( \cdots \dynamics(\ \ \ F(\statee, t_i\ ;\param^{(j)}),\ \action_{1+t_i}\ ), \action_{2+t_i}), \cdots), \action_{t_{i+1}-2}),\ \action_{t_{i+1}-1}),
    $$ 
    we have  $\samplednospotj(0) = \statee_0$, and for all $i \in \left\{0,1, \cdots, N-1 \right\}$, we have,
    $$
    \samplednospotj(i+1) = \dynamics_{i+1}^{(j)}\left(\samplednospotj(i)\right).
    $$
\end{definition}
\begin{remark}
The sub-trajectory $\subnospotj$ with parameters $\param^{(j)}$ can also be recursively defined as:
$$
\begin{aligned}
    &\subnospotj(i+1) = \\
    &F\left(\cdots\left(\ \ \ F\left(F\left(\ \subnospotj(i),t_i \ ;\param^{(j)}\ \right),t_i+1\ ;\param^{(j)}\right) \cdots\right), t_{i+1}-1\ ;\param^{(j)}\right).
\end{aligned}
$$
Notice that the parameters $\param^{(j)}$ are referenced multiple times while in $\samplednospotj$ only once.
\end{remark}
Figure~\ref{fig:sampling} \navidg{presents} Definition \ref{def:straj} through visualization. This definition replaces the set of selected nodes - on a randomly selected time-step - with its pre-computed evaluation. \navidg{This set of nodes are indeed a controller unit on the time-steps sampled to apply dropout}\footnote{The set of sampled time-steps for dropout is in fact the set-difference between $\mathcal{K}$ and $\mathcal{T}$, where $\mathcal{T}$ is the set of sampled times steps that is generated to define the sampled trajectory.}. \navidg{ Excluding the time steps with fixed actions, we then name the set of states on the remaining timesteps -  as the {\em sampled trajectory}, and we denote it as $\samplednospotj$.}
\begin{example}\label{ex:sampledtraj}
    Let the state and action at the time $\timeid$ be $\mathbf{x}_\timeid \in \mathbb{R}$  and $\text{a}_\timeid \in \mathbb{R}$, respectively. The feedback controller is $\text{a}_\timeid = \policy_{\param}(\mathbf{x}_\timeid, \timeid), \param \in \mathbb{R}^3$ and the dynamics is also $\mathbf{x}_{\timeid+1} = \dynamics( \mathbf{x}_\timeid , \text{a}_\timeid),\  \mathbf{x}_0 = 1.15$. Let’s also assume a trajectory of horizon $9$ over time-domain (i.e., $\mathcal{K} = \{ i \mid 0 \le i \le 9\}$) with a trajectory  $\traj[\mathbf{x}_0\ ;\param] = \mathbf{x}_0, \mathbf{x}_1, \mathbf{x}_2, \mathbf{x}_3, \mathbf{x}_4, \mathbf{x}_5, \mathbf{x}_6, \mathbf{x}_7, \mathbf{x}_8, \mathbf{x}_9 $. Suppose, we are in the gradient step $j=42$, and in this iteration, we want to generate a sampled trajectory with $N=3$ time-steps, where, $\mathcal{T} = \left\{ 0, t_1 = 1, t_2 = 3, t_3 = 6\right\}$. The control parameters at this gradient step are also $\param^{(42)} = \left[ 1.2,2.31,-0.92\right]$ that results in the control sequence $\textbf{a}= 0,0.1,0.2,0.3,0.4,0.5,0.6,0.7,0.8$. Given this information, we define the sampled trajectory as the sequence $\mathsf{smpl}\left(\traj[\mathbf{x}_0\ ;\param^{(42)}], \mathcal{T}\right)  = \mathbf{x}_0, \tilde{\mathbf{x}}_1, \tilde{\mathbf{x}}_3, \tilde{\mathbf{x}}_6$, where, \\
    \begin{minipage}[t]{0.68\linewidth}
    \vspace{-15mm}
    \[
    \begin{aligned}
    &\tilde{\mathbf{x}}_1 = \dynamics_1^{(42)}(\mathbf{x}_0) = F(\mathbf{x}_0, 0\ ;\param^{(42)}),\\
    &\tilde{\mathbf{x}}_3 = \dynamics_2^{(42)}(\tilde{\mathbf{x}}_1) = \dynamics(\ F(\tilde{\mathbf{x}}_1, 1\ ;\param^{(42)}),\  0.2),\\
    &\tilde{\mathbf{x}}_6 = \dynamics_3^{(42)}(\tilde{\mathbf{x}}_3) = \dynamics(\ \dynamics(\ F(\tilde{\mathbf{x}}_3, 3\ ;\param^{(42)}),\  0.4), \ 0.5).
    \end{aligned}\ \implies
    \]
    \end{minipage}%
    \hfill
    \begin{minipage}{0.3\linewidth}
    \includegraphics[width=\linewidth]{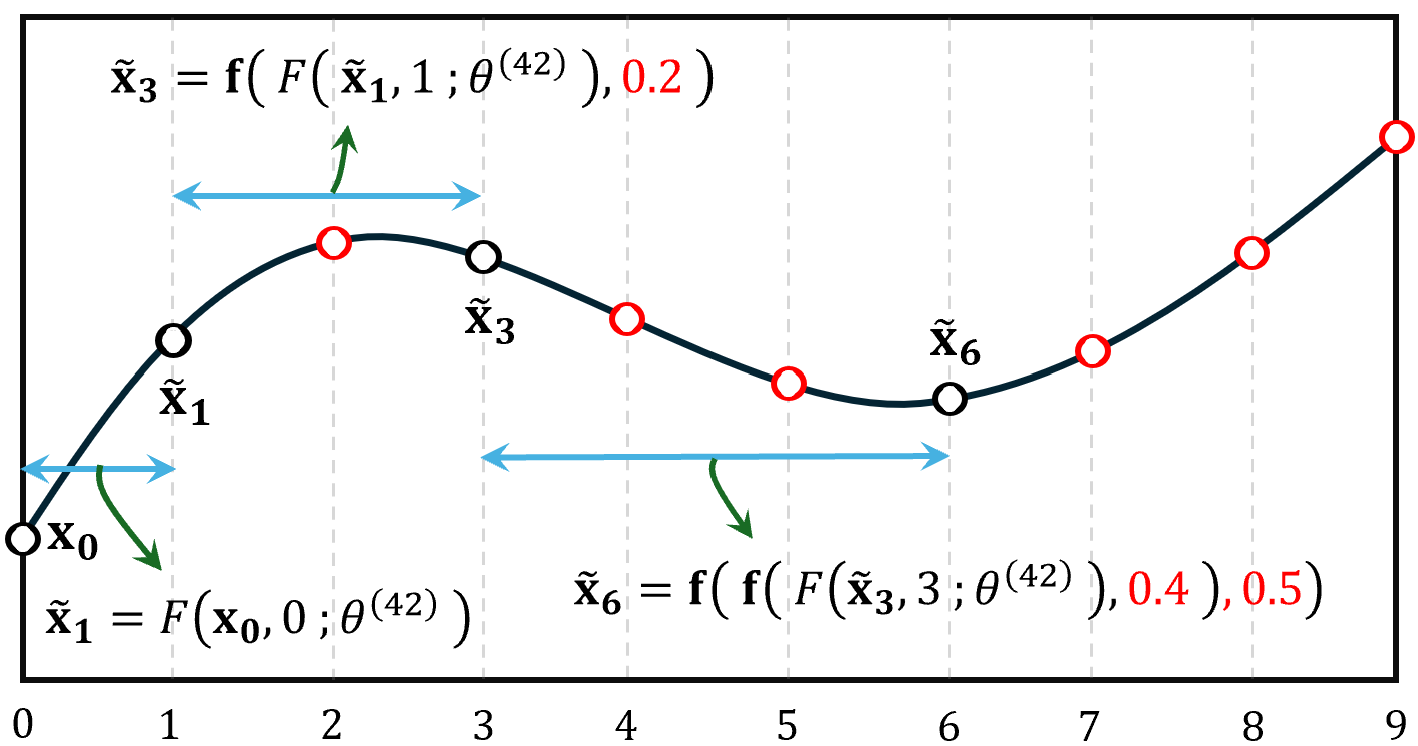}
    \end{minipage}
    where the constants $0.2, 0.4,0.5$ are the $3^{\text{rd}}, 5^{\text{th}}$, and $6^{\text{th}}$ elements in the pre-evaluated control sequence $\textbf{a}$, respectively.
\end{example}

\begin{wrapfigure}[14]{r}{0.55\textwidth}
    \vspace{-7.5mm}
    \hspace{-4mm}
    \includegraphics[width=1.05\linewidth]{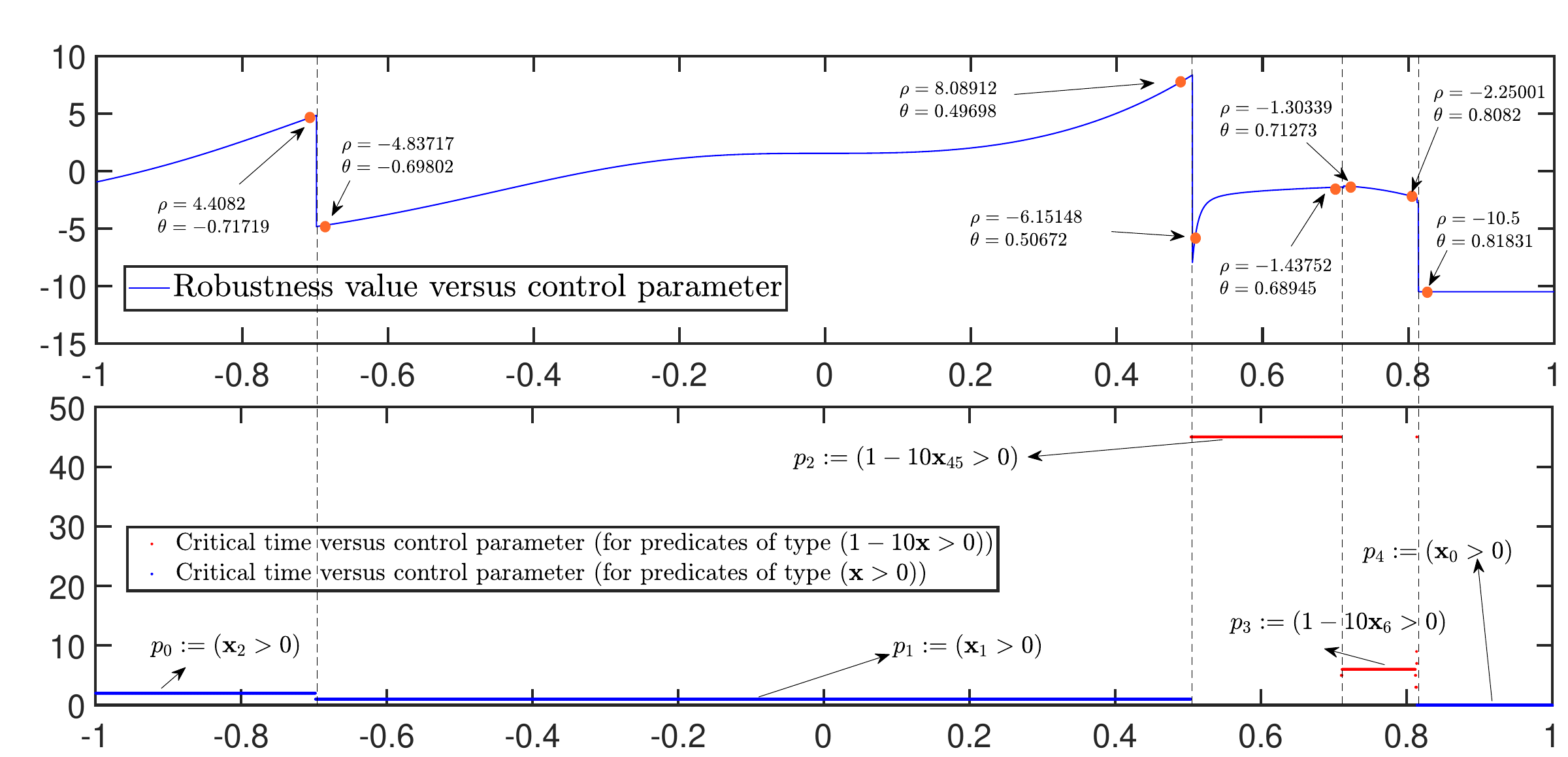}
    \captionsetup{skip = 0mm}
    \caption{this figure shows a common challenge in using critical predicate for control synthesis. This figure presents the robustness as a piece-wise differentiable function of control parameter $\param$ (with resolution, $0.00001$), where each differentiable segment represent a distinct critical predicate.}
    \label{fig:ex3}
\end{wrapfigure}
\subsection{Including the Critical Predicate in Time Sampling}\label{subsec:critic}
While it is possible to select random time-points to use in the gradient computation, in our preliminary results, exploiting the structure of the given DT-STL formula -- specifically identifying and \navidg{ using {\em critical predicates} \cite{abbas2013computing}} -- gives better results. \navidg{Proposition 3.1 in \cite{abbas2013computing} introduces the notion of critical predicate. Here, we also provide this definition as follows:} 

\begin{definition}[Critical Predicate]
As the robustness degree of DT-STL is an expression consisting of $\min$ and $\max$ of robustness values of predicates at 
different times, the robustness degree is consistently  equivalent to the robustness of one of the predicates $h(\cdot)$ at a specific time. This specific predicate $h^*>0$ is called the critical predicate, and this specific time $\timeid^*$ is called the critical time.  
\end{definition}

\begin{example}\label{ex:critic}
    We again consider Example \ref{ex:1} to clarify the notion of critical predicate. In this example, we have $4$ predicates of a unique type, e.g. $h(\mathbf{x}_k) = \mathbf{x}_k > 0$. Thus, the \navidddd{robustness values} of the predicate $h(\mathbf{x}) > 0$ at time points $0,1,2,3$ are respectively $\mathbf{x}_0,\mathbf{x}_1,\mathbf{x}_2,\mathbf{x}_3$. Assume the trajectory is $\traj[\mathbf{x}_0\ ;\param] = \left[ \mathbf{x}_0=1,\ \mathbf{x}_1 =2,\ \mathbf{x}_2=3,\ \mathbf{x}_3=1.5 \right]$. Since the robustness function is defined as $\rob(\varphi,0) = \max\left( h(\mathbf{x}_0),\ h(\mathbf{x}_1),\ h(\mathbf{x}_2),\ h(\mathbf{x}_3) \right)$, the robustness \navidddd{value} is equivalent to $h(\mathbf{x}_2)$. Thus, we can conclude, the critical predicate is $h^* = h(\mathbf{x}_2)> 0$ and the critical time is $\timeid^* = 2$. 
\end{example}
The critical predicate and critical time of a DT-STL formula can be computed using the same algorithm used to compute the robustness value for a given DT-STL formula.
\navidg{ This algorithm is implemented in the S-Taliro tool \cite{annpureddy2011s}}.

\subsection{Safe Re-Smoothing} \label{subsec:saferesmoothing}

A difficulty in using critical predicates is that a change in controller parameter values may change the system trajectory, which may in turn change the predicate that is critical in the robustness computation. Specifically, if the critical predicate in one gradient step is different from the critical predicate in the subsequent gradient step, our gradient ascent strategy may fail to improve the robustness value, as the generated gradient \navidg{in this gradient step is local}.

\begin{example}\label{ex:piecewise}
   To clarify this with an example, we present a specific scenario in Figure~\ref{fig:ex3}. This figure shows the robustness value as a non-differentiable function of control parameters, that is a piece-wise differentiable relation where every differentiable segment represents a specific critical predicate. The system dynamics is $\mathbf{x}_{\timeid+1} = 0.8 \mathbf{x}_{\timeid}^{1.2} - \mathbf{e}^{-4u_\timeid\sin(u_\timeid)^2}$, where the system starts from $\mathbf{x}_0 = 1.15$ and the controller is $u_\timeid = \tanh(\param \mathbf{x}_\timeid)$. The robustness is plotted based on control parameter $-1 \leq \param \leq 1$ and is corresponding to the formula $\Phi = \ev_{[0,45] }\left( \alw_{[0,5] } \left(\mathbf{x}>0\right) \right) \wedge \alw_{[0,50] }\left(1-10\mathbf{x}>0\right)$. Assume the training process is in the $15^{\text{th}}$ gradient step of back-propagation with $\param = \param^{(15)}=0.49698$ where the critical predicate for this control parameter is denoted by $p_1:= (\mathbf{x}_1>0)$. The gradient generated from the critical predicate $p_1$ suggests increasing the value of $\param$, \navidg{which should result in $\theta = \theta^{(16)}= 0.50672$. However, applying the gradient would move the parameter value to a region of parameter space} where the critical predicate is $p_2:= (1-10\mathbf{x}_{45}>0)$. \navidg{In this case, the gradient generated from the critical predicate \(p_1\) is local to this gradient step, as the critical predicate shifts from \(p_1\) to \(p_2\). Our approach in this scenario is to first reduce the learning rate. If this does not lead to an increase in the robustness value, we then transition to smooth semantics, which takes all predicates into account.} The scenario proposed in this figure shows this local gradient may result in a drastic drop in the robustness value from $8.09$ to $-6.15$. Therefore, the gradient of critical predicate is useful, only if the gradient step preserves the critical predicate.
\end{example}
\begin{figure*}
  \begin{minipage}{0.49\textwidth}
    \begin{mdframed}
    \includegraphics[width=\linewidth]{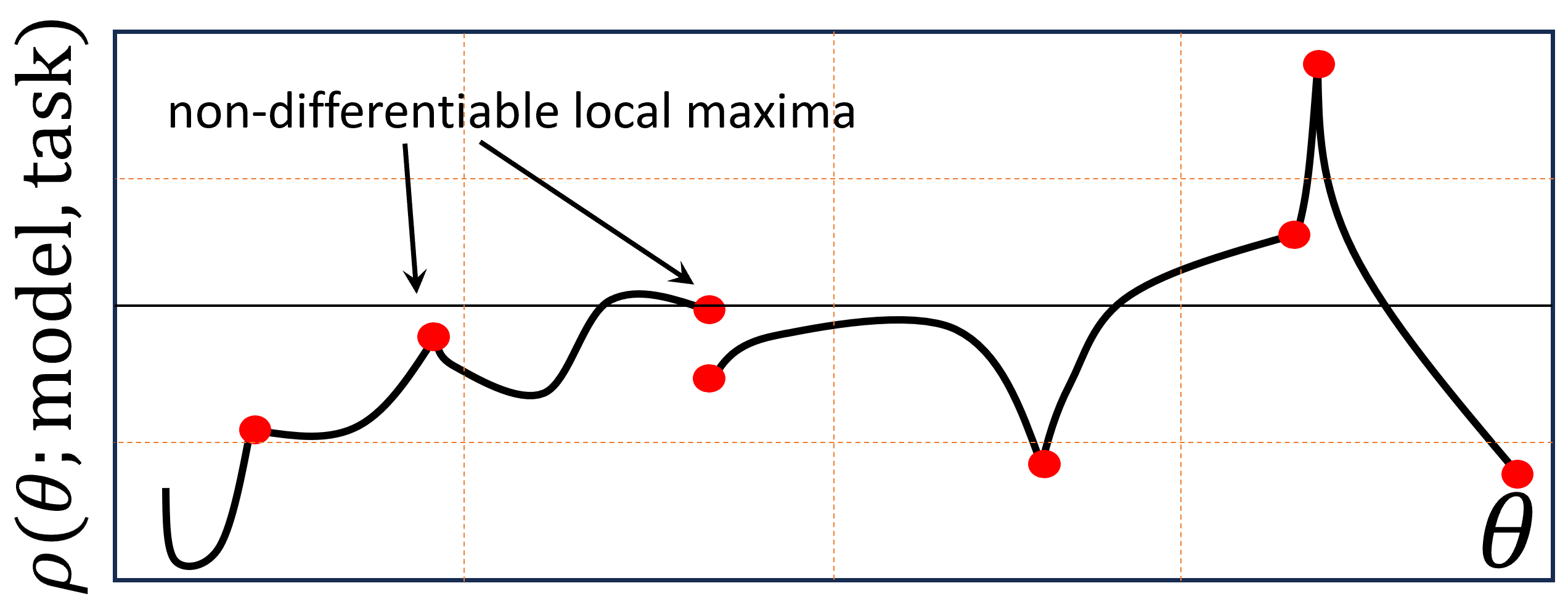}
    \caption{\navidddd{This figure shows} an example for the relation between control parameters and the resulting robustness as a \navidddd{piece-wise differentiable} function. Assuming a fixed initial state, every control parameter is corresponding to a simulated trajectory, and that trajectory represents a robustness value. This robustness value is equal to the quantitative semantics for the critical predicate. \navidddd{Within each differentiable segment in this plot, the control parameters yield trajectories associated with a unique critical predicate.}}\label{fig:nsmoothpoint}
    \end{mdframed}
    
  \end{minipage}
  \hfill
  \begin{minipage}{0.49\textwidth}
    \begin{mdframed}
    \hspace{-4mm}
    \includegraphics[width=1.1\linewidth]{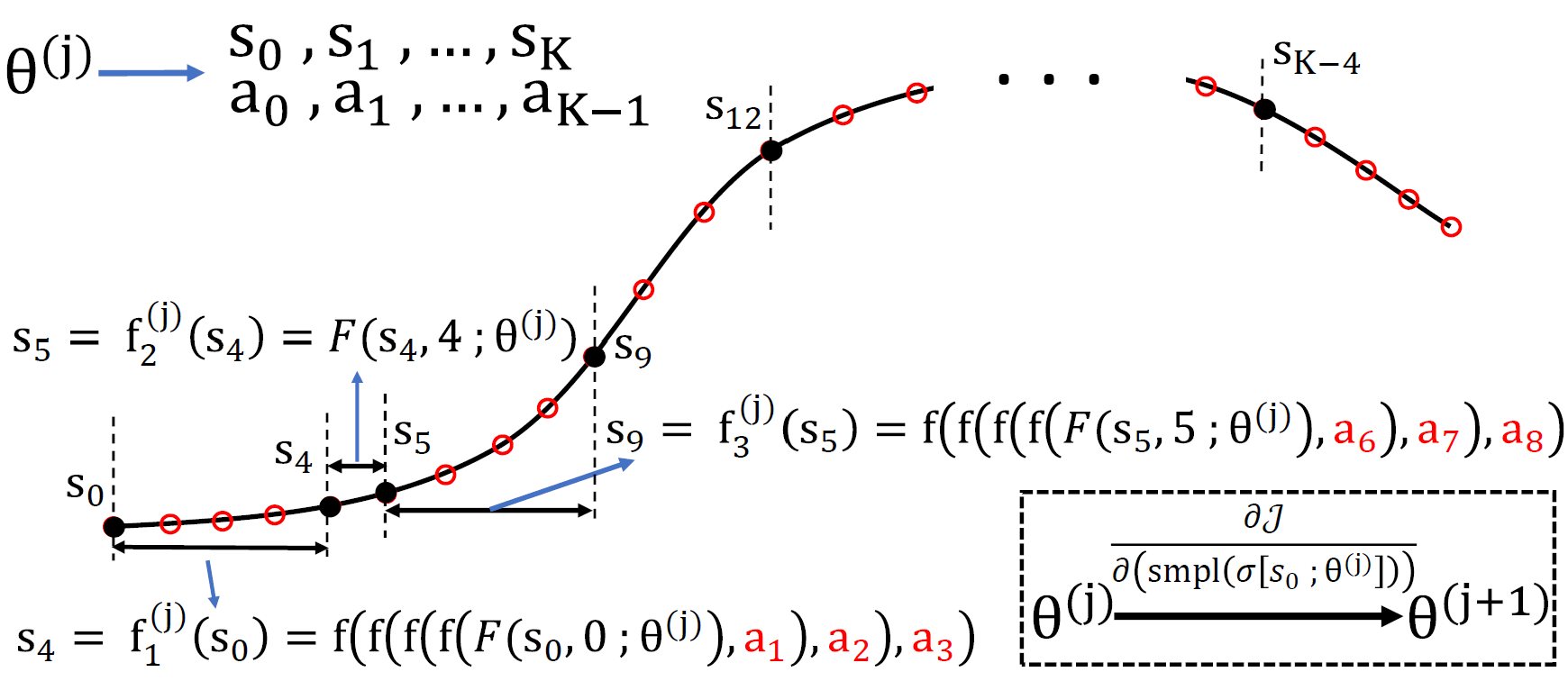}
    \end{mdframed}
    \caption{This figure depicts the sampling-based gradient computation. In our approach, we freeze the controller at some time-points, while at others we assume the controller to be a function of its parameters that can vary in this iteration of back-propagation process. The actions that are fixed are highlighted in red, whereas the dependent actions are denoted in black. The red circles represent the time-steps where the controller is frozen.}
    \label{fig:sampling} 
  \end{minipage}
\end{figure*}

Given a predefined specification $\varphi$, a fixed initial state, differentiable controller with parameter $\param$, and a differentiable model, the robustness value is a piece-wise differentiable function of control parameter, where each differentiable segment represents a unique critical predicate (see Figure~\ref{fig:nsmoothpoint}). However, \navidg{the Adam algorithm\footnote{In this paper, we utilize MATLAB's $\mathsf{adamupdate()}$ library, \url{https://www.mathworks.com/help/deeplearning/ref/adamupdate.html}} } assumes a differentiable objective function. Therefore, we utilize the critical predicate as the objective function when we are in the differentiable segments, and we replace it with the smooth semantics of DT-STL robustness, $\stltolb$, at the non-differentiable local maxima where the critical predicate is updated. We refer to this shift between critical predicate and smooth semantics as safe re-smoothing. However, it is practically impossible to accurately detect the non-differentiable local maxima, thus we take a more conservative approach and we instead, utilize $\stltolb$ at every gradient step when the critical predicate technique is unable to improve the robustness.

\subsection{Computing the Sampled Gradient} \label{subsec:approxcompute}
We now explain how we compute an approximation of the gradient of original trajectory (that we call the {\em original gradient}). We call the approximate gradient from our sampling technique as the \textit{sampled gradient}.
In the back-propagation algorithm - at a given gradient step $j$ and with
control parameter $\param^{(j)}$ - we wish to compute the sampled gradient
$[\partial {\mathcal{J}}/ \partial {\param^{(j)}}]_{\text{sampled}}$. The
objective function $\mathcal{J}$ in our training algorithm can be either the
robustness for critical predicate or the smooth semantics for the robustness of
trajectory, $\tilde{\rob}$. The former is defined over a single trajectory
state, (i.e., at critical time) while the latter is defined over the entire
trajectory. In response, we propose two different approaches for trajectory
sampling for each objective function. 


\begin{enumerate}[wide, nosep, labelwidth=!, labelindent=0pt]
\item In case the objective function $\mathcal{J}$ is the robustness for critical predicate, it is only a function of the trajectory state $\statee_{\timeid^*}$. Thus, we sample the time-steps as, $\mathcal{T} = \left\{ 0,t_1 , t_2 , \cdots , t_N \right\}, \ t_N = \timeid^*$ to generate a sampled trajectory $\samplednospotj$ that ends in critical time. 
We utilize this sampled trajectory to compute the sampled gradient. The original gradient regarding the critical predicate can be formulated as, 
$\partial \mathcal{J}/\partial \param = \left(\partial \mathcal{J}/\partial \statee_{\timeid^*}\right) \left( \partial \statee_{\timeid^*}/\partial \param \right)$, where $\statee_{\timeid^*}=\subnospotj(N)$.
However, we define $\mathcal{J}$ on our sampled trajectory and propose the sampled gradient as, 
\[
\left[\frac{\partial \mathcal{J}}{\partial \param}\right]_{\text{sampled}} = \left(\frac{\partial \mathcal{J}}{\partial \samplednospotj(N)}\right)\left(\frac{\partial \samplednospotj(N)}{\partial \param}\right).
\]
\item In case the objective function is the smooth semantics for the robustness $\stltolb$, it is a function of all the trajectory states. In this case, we consequently segment the trajectory into $M$ subsets, by random time sampling as, \navidg{$  \mathcal{T}^q = \left\{0, t_1^q, t_2^q, \cdots, t_N^q \right\} \subseteq \mathcal{K}, \  q\in \left\{1,\cdots,M\right\}$} (See Example \ref{ex:sampleall}), where,
\begin{equation} \label{eq:samplerules}
(\forall q,q' \in  \left\{ 1,\cdots,M \right\} :\mathcal{T}^{q} \cap \mathcal{T}^{q'}= \left\{ 0 \right\}) \wedge (\mathcal{K} = \bigcup_{q\in \left\{1,\cdots,M\right\}} \mathcal{T}^q).
\vspace{-3mm}
\end{equation}
Let's assume the sub-trajectories $\subspotqj= \statee_0, \statee_{t_1^q}, \cdots, \statee_{t_N^q}$ and their corresponding sampled trajectories as $\sampledspotqj$. As the sampled time-steps $\mathcal{T}^q, q\in \left\{ 1,\cdots,M\right\}$ have no time-step in common other than $0$ and their union covers the horizon $\mathcal{K}$, we can reformulate the original gradient ($\partial{\mathcal{J}}/\partial \param= \sum_{\timeid=1}^\horizon (\partial{\mathcal{J}}/\partial \statee_\timeid)(\partial{\statee_\timeid}/\partial \param)$) as: 
$$
\frac{\partial \mathcal{J}} {\partial \param} = \sum_{q=1}^M \left( \frac{ \partial \mathcal{J} } { \partial \subspotqj } \right)\left( \frac{ \partial \subspotqj }  {\partial \param } \right).
$$
However, in our training process to compute the sampled gradient, we relax the sub-trajectories $ \subspotqj, q\in \left\{1,\cdots,M\right\}$ with their corresponding sampled trajectories $\sampledspotqj$. In other words, 

$$
\left[\frac{\partial \mathcal{J}}{\partial \param}\right]_{\text{sampled}} = \sum_{q=1}^M \left( \frac{\partial \mathcal{J}}{\partial \sampledspotqj} \right)\left( \frac{\partial \sampledspotqj}{\partial \param}  \right).
$$
\end{enumerate}

\begin{remark}
Unlike ${\partial \statee_{\timeid^*}}/{\partial \param}$ and ${\partial \subspotqj }/{\partial \param}, q \in  \left\{1,\cdots,M\right\}$ that are prone to vanish/explode problem, the alternatives, ${\partial \samplednospot(N)}/{\partial \param}$, and ${\partial \sampledspotq}/{\partial \param}, q \in \left\{ 1,\cdots,M\right\}$,  can be computed efficiently\footnote{The efficiency results from the control parameters $\param$ repeating in fewer time-steps over the trajectory, as most of them are fixed.}.   
\end{remark}
\begin{example}\label{ex:sampleall}
    Here, we propose an example to show our methodology to generate sampled trajectories when $\mathcal{J}= \tilde{\rho}$. We again consider the Example \ref{ex:sampledtraj}, but we sample the trajectory with $M=3$ sets of sampled time-steps $\mathcal{T}^1 = \left\{0,2,4,9 \right\},\ \mathcal{T}^2 = \left\{0,5,7,8 \right\}$ and $\mathcal{T}^3 = \left\{0,1,3,6\right\}$. Here, the time-steps are sampled such that their intersection is $\left\{ 0 \right\}$ and their union is $\mathcal{K}$. The resulting sampled trajectory for $\mathcal{T}^1$ is $\mathsf{smpl}\left(\traj[\mathbf{x}_0\ ;\param^{(42)}], \mathcal{T}^1\right) = \mathbf{x}_0, \tilde{\mathbf{x}}_2, \tilde{\mathbf{x}}_4, \tilde{\mathbf{x}}_9$, where, 
    \begin{minipage}[t]{0.68\linewidth}
    \vspace{-12mm}
    \[
    \resizebox{0.95\linewidth}{!}{$
    \begin{aligned}
    &\tilde{\mathbf{x}}_2 = \dynamics_1^{(42)}(\mathbf{x}_0) = \dynamics(\ F(\mathbf{x}_0, 0\ ;\param^{(42)}),\ 0.1),\\
    &\tilde{\mathbf{x}}_4 = \dynamics_2^{(42)}(\tilde{\mathbf{x}}_2) = \dynamics(\ F(\tilde{\mathbf{x}}_2, 2\ ;\param^{(42)}),\  0.3),\\
    &\tilde{\mathbf{x}}_9 = \dynamics_3^{(42)}(\tilde{\mathbf{x}}_4) = \dynamics(\ \dynamics(\ \dynamics(\ \dynamics(\ F(\tilde{\mathbf{x}}_4, 4\ ;\param^{(42)}),\  0.5), \ 0.6), \ 0.7), \ 0.8),
    \end{aligned}\ \implies
    $}
    \]
    \end{minipage}%
    \hfill
    \begin{minipage}{0.3\linewidth}
    \includegraphics[width=\linewidth]{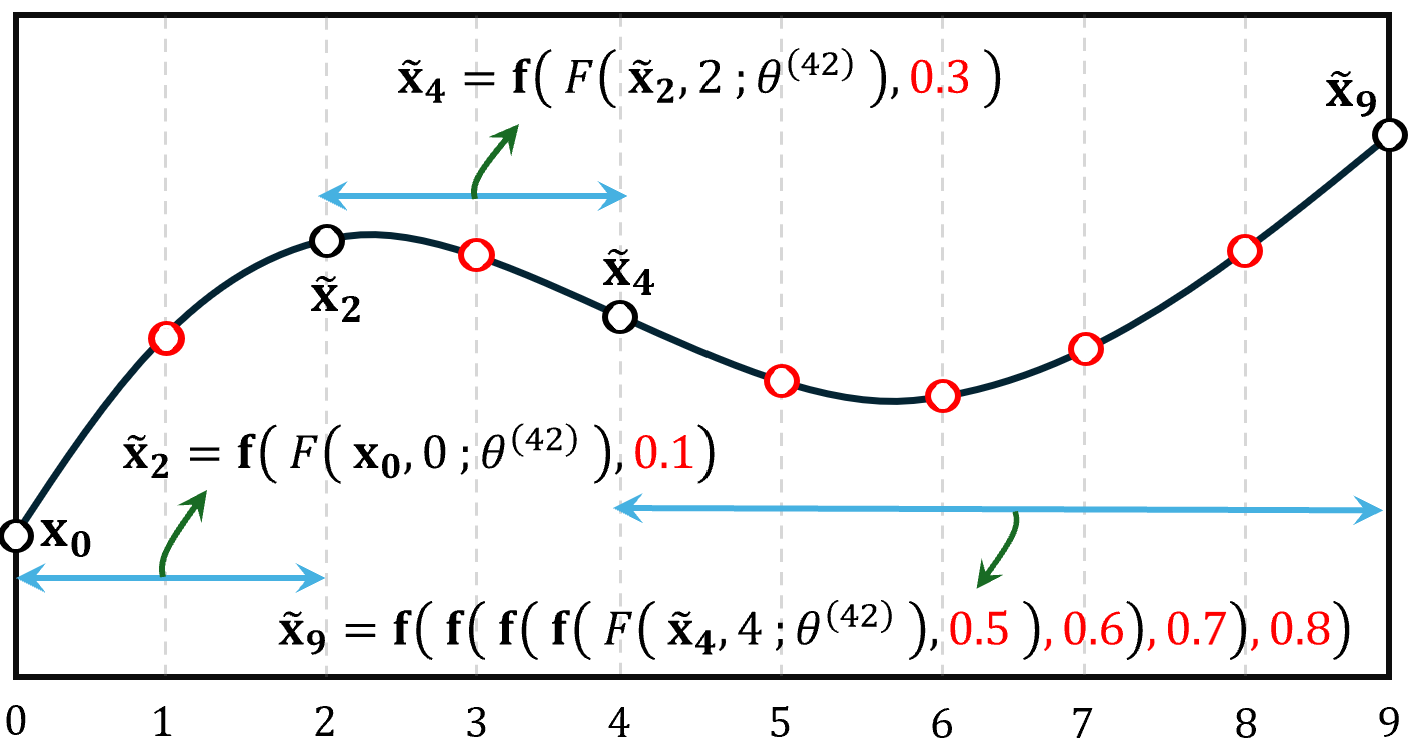}
    \end{minipage}
    
    \noindent and the resulting sampled trajectory for $\mathcal{T}^2$ is $\mathsf{smpl}\left(\traj[\mathbf{x}_0\ ;\param^{(42)}], \mathcal{T}^2\right) = \mathbf{x}_0, \tilde{\mathbf{x}}_5, \tilde{\mathbf{x}}_7, \tilde{\mathbf{x}}_8$, where,\\
    \begin{minipage}[t]{0.68\linewidth}
    \vspace{-12mm}
    \[
    \resizebox{0.95\linewidth}{!}{$
     \begin{aligned}
     &\tilde{\mathbf{x}}_5 = \dynamics_1^{(42)}(\mathbf{x}_0) = \dynamics(\ \dynamics(\ \dynamics(\ \dynamics(\ F(\mathbf{x}_0, 0\ ;\param^{(42)}),\  0.1), \ 0.2), \ 0.3), \ 0.4),\\
     &\tilde{\mathbf{x}}_7 = \dynamics_2^{(42)}(\tilde{\mathbf{x}}_5) = \dynamics(\ F(\tilde{\mathbf{x}}_5, 5\ ;\param^{(42)}),\ 0.6),\\
     &\tilde{\mathbf{x}}_8 = \dynamics_3^{(42)}(\tilde{\mathbf{x}}_7) = F(\tilde{\mathbf{x}}_7, 7\ ;\param^{(42)}),
     \end{aligned}\ \implies
     $}
    \]
    \end{minipage}%
    \hfill
    \begin{minipage}{0.3\linewidth}
    \includegraphics[width=\linewidth]{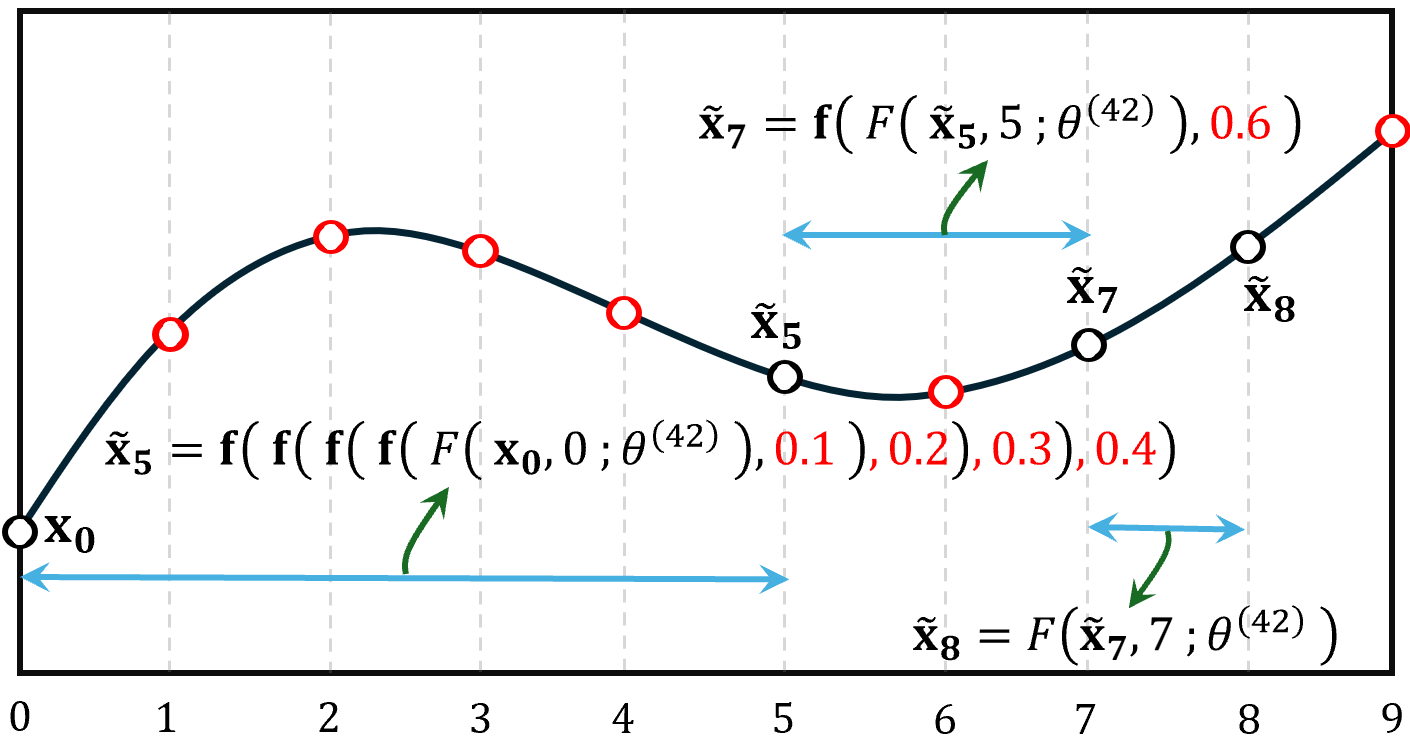}
    \end{minipage}
    
    \noindent and finally, the resulting sampled trajectory for $\mathcal{T}^3$ is $\mathsf{smpl}\left(\traj[\mathbf{x}_0\ ;\param^{(42)}], \mathcal{T}^3\right) =  \mathbf{x}_0, \tilde{\mathbf{x}}_1, \tilde{\mathbf{x}}_3, \tilde{\mathbf{x}}_6$ that has been previously explained in Example~\ref{ex:sampledtraj}.
    We emphasize that the introduced sampled trajectories are exclusively generated for gradient step $j=42$ and we perform a new random sampling for the next iteration.\vspace{-3mm}
\end{example}
\RestyleAlgo{ruled}
\DontPrintSemicolon
\begin{algorithm*}[t]
     \tiny
     $\textbf{Input:\ }  \epsilon,\ M,\ N,\  N_1,\  N_2,\ \param^{(0)},\  \varphi,\ \bar{\rob},\  \sampledinits,\  j=0$ \;
     \While{$\rob^\varphi(\traj[\statee_0\ ;\param^{(j)}]) \leq \bar{\rob}$ \label{line:terminate}}{
        $\statee_0 \gets$ $\mathtt{Sample\ from\ }  \sampledinits$\qquad
        $\mathtt{use\_smooth} \gets  \mathsf{False} \quad  j \gets j+1$   \label{line:sampleint}\;
        \If{$\mathtt{use\_smooth} = \mathsf{False}$}{
        $\param_1,\ \param_2  \gets \param^{(j)}$  \tcp*{\textcolor{blue}{$\param_1 \& \param_2\ \mathtt{are\ candidates\ for\ parameter\ update\ via\ critical\ predicate\ and\ waypoint.}$}}
        \tcp{\textcolor{blue}{$\mathtt{The\ following\ loop\ updates\ \param_1\ and\ \param_2\  via\ cumulation\ of\ N_1\ sampled\  gradients}$}}
        \For{$i \gets 1, \cdots , N_1$}{
            $\traj[\statee_0\ ;{\param_1}],\ \traj[\statee_0\ ;{\param_2}] \gets$ $\mathtt{Simulate\ the\ trajectory\ via\ \param_1,\ \param_2,\   and\ \statee_0 }$\;
            $\timeid^*, h^*(\statee_{\timeid^*}) \gets \mathtt{\ obtain\ the\ critical\ time\ } \mathtt{\ and\ the\ critical\ predicate\ } $\;
            $\mathcal{T}^1,\  \samplednospotf\ \hspace{-0.5mm} \gets \hspace{-0.5mm}$ $ \mathtt{ sample\ set\ of\ }\mathtt{\ time\ steps\ } \mathcal{T}^1=\left\{0,t_1,..,t_N=\timeid^* \right\} \mathtt{\ and\ its\ sampled\ trajectory }$\;
            $\mathcal{T}^2,\  \samplednospotfwp   \hspace{-0.5mm} \gets \hspace{-0.5mm}$ $ \mathtt{ sample\ set\ of\ }\mathtt{\ time\ steps\ } \mathcal{T}^2=\left\{0,t_1,..,t_N \right\} \mathtt{\ and\ its\ sampled\ trajectory }$\;
            $\mathcal{J} \gets h^*\left(\samplednospotf(N)\right)$\qquad  $d_1 \gets [\partial \mathcal{J}/\partial \param]_{\text{sampled}}$ \qquad $\param_{1} \gets
                \param_1+\adam(d_1/N_1)$ \nllabel{algoline:stlupdate} \;
            $\mathcal{J} \gets \mathcal{J}^{\text{wp}}\left(\samplednospotfwp\right)$\qquad  $d_2 \gets [\partial \mathcal{J}/\partial \param]_{\text{sampled}}$ \qquad $\param_{2} \gets
                \param_2+\adam(d_2/N_1)$ \nllabel{algoline:stlupdatewp} \;
            \nllabel{algoline:firstif}
        }
         \tcp{\textcolor{blue}{$\mathtt{Update\ the\ control\ parameter\ with\ \param_2\ if\ it\ increases\ the\ robustness\  value}$}}
         \tcp{\textcolor{blue}{$\mathtt{Otherwise\ update\ the\ control\ parameter\ with\ \param_1\ if\ it\ increases\ the\ robustness\  value}$}}
         \tcp{\textcolor{blue}{$\mathtt{Otherwise,\  check\ for\ non-differentiable\ local\ maximum}$}}
         \lIf{$\rob^\varphi(\traj[\statee_0\ ;{\param_{2}}]) \ge 
              \rob^\varphi(\traj[\statee_0\ ;{\param^{(j)}}])$}{
              $\param^{(j+1)} \gets
                  \param_{2}$
         }
         \lElseIf{$\rob^\varphi(\traj[\statee_0\ ;{\param_{1}}]) \ge 
              \rob^\varphi(\traj[\statee_0\ ;{\param^{(j)}}])$}{
              $\param^{(j+1)} \gets
                  \param_{1}$
         }
         \Else{
                 $\ell  \gets 1 \qquad  \mathsf{update} \gets \mathsf{True}$\;
                 \While{$\mathsf{update}\  \&\  (\mathtt{use\_smooth}\! =\! \mathsf{False})$ }{
                    $\ell \gets \ell/2$\quad
                    $\hat{\param} \gets \param^{(j)} + \ell(\param_1-\param^{(j)})$ \label{line:nondifdetect1} \tcp*{\textcolor{blue}{$\mathtt{ Keep\  the\  gradient\ direction\ \& \  reduce\  the\  learning\  rate}$}}
                    \tcp{\textcolor{blue}{$\mathtt{Update\ the\ control\ parameter\ with\ }\hat{\param}\mathtt{\ if\ it\ increases\ the\ robustness\  value}$}}
                    \lIf{$\rob(\varphi,\traj[\statee_0\ ;{\hat{\param}}],0)    \ge \rob^\varphi(\traj[\statee_0\ ;{\param^{(j)}}])$}{
                         $\left[ \param^{(j+1)} \gets 
                             \hat{\param} \qquad \mathsf{update} \gets \mathsf{False}\right]$ \label{line:nondifdetect2}
                    }
                    \ElseIf{$\ell < \epsilon$}{
                          $\mathtt{use\_smooth} \gets \mathsf{True} $  \label{line:updatesmooth} \tcp*{\textcolor{blue}{$\mathtt{\ swap\ the\ objective\ with\ }\tilde{\rob}\mathtt{\ if\ } \ell < \epsilon$}}
                    }
                }
         }
     }
     \If{$\mathtt{use\_smooth} = \mathsf{True} $}{
                \hspace{-1mm}$\param_3 \gets \param^{(j)}$ \tcp*{\textcolor{blue}{$\param_3\ \mathtt{is\ the\ candidate\ for\ parameter\ update\ via\ smooth\ semantic\ } \tilde{\rob}$}}
                \tcp{\textcolor{blue}{\hspace{-1mm} $\mathtt{The\ following\ loop\ updates\ \param_3\ via\ cumulation\ of\ N_2\ sampled\  gradients}$}}
                \hspace{-1mm}\For{$i \gets 1, \cdots , N_2$}{
                $\hspace{-2mm}\mathcal{T}^q,  \sampledspotqs , q\in 1,\cdots,M  \gets \hspace{-1mm} \mathtt{ Make\ } M \mathtt{\ sets\ of\ sampled\  time\ steps\ from\ } Eq.~\eqref{eq:samplerules} \mathtt{\ \&\ their\  sampled\ trajectories }$\;
                $\hspace{-2mm} \mathcal{J} \gets \tilde{\rob} \qquad d_3 \gets [\partial \mathcal{J}/\partial \param]_{\text{sampled}}$\qquad
                $\param_3 \gets
                \param_3+\adam(d_3/N_2)$\;
            }
            \hspace{-1mm}$\param^{(j+1)} \gets \param_3$\;
        }
    }
\caption{Gradient sampling and training the controller for long horizon tasks.}
\label{algo:training}
\end{algorithm*}
\begin{remark}
    \navid{At the start of the training process, we can envision a desired path for the model to track. Tracking this path may not be sufficient to satisfy the temporal specification, but its availability is still valuable information, which its inclusion to the training process can expedite it. Therefore, we also utilize a desired path to generate a convex and efficient waypoint function (denoted by $\mathcal{J}^{\text{wp}}\left(\traj\left[\statee_0\ ;\param\right]\right)$) for our training process. However, Algorithm~\ref{algo:training} performs effectively even without the waypoint function. Section~\ref{apdx:analysis} explores this aspect using a numerical example. Nonetheless, integrating a waypoint function enhances the efficiency of the training process.}
\end{remark}
We finally present our overall training procedure in Algorithm~\ref{algo:training}. Here, we use $\rob^\varphi(\traj[\statee_0\
;\param])$ as shorthand for the non-smooth robustness degree of
$\traj[\statee_0\ ;\param]$ w.r.t. $\varphi$ at time $0$, \navidg{i.e., $\rob(\varphi, \traj[\statee_0\
;\param], 0)$ }. 
We terminate
the algorithm in line~\ref{line:terminate} if the robustness is greater than a
pre-specified threshold $\bar{\rho}>0$. We also evaluate the performance of
the algorithm through challenging case studies.
During each iteration of this algorithm, we compute the robustness value for an
initial state $\statee_0$ selected from the pre-sampled set of initial states
$\sampledinits$ in line~\ref{line:sampleint}. This selection can be
either random, or the initial state with the lowest robustness value in the set
$\sampledinits$. 
The Boolean parameter $\mathsf{use\_smooth}$ is provided to toggle the objective between robustness of the critical predicate and the smooth robustness for the DT-STL formula. We initialize this parameter $\mathsf{use\_smooth}$ in line~\ref{line:sampleint} to be $\mathsf{False}$ and further update it to $\mathsf{True}$ in line~\ref{line:updatesmooth}, in case the gradient from critical predicate is unable to increase the robustness. 
\navid{The lines~\ref{line:nondifdetect1},\ref{line:nondifdetect2} and \ref{line:updatesmooth} aim to improve the detection of non-differentiable local maxima by employing a more accurate approach. This involves maintaining the direction of the gradient generated with the critical predicate, and exponentially reducing the learning rate until a small threshold $\epsilon$ is reached. If, even with an infinitesimal learning rate, this gradient fails to increase the robustness, it suggests a high likelihood of being in a non-differentiable local maximum.}

\section{Experimental Evaluation}
\label{sec:experiments}

In this section, we evaluate the performance of our proposed methodology. We executed all experiments for training with Algorithm~\ref{algo:training} using our MATLAB toolbox\footnote{The source code for the experiments is publicly available from \url{https://github.com/Navidhashemicodes/STL_dropout}}. These experiments were carried out on a laptop PC equipped with a Core i9 CPU. In all experiments performed using Algorithm~\ref{algo:training}, we utilize $\lbfortl$ as the smooth semantics. We also present an experiment in Section~\ref{sec:feedback} to empirically demonstrate that NN feedback controllers provide robustness to noise compared to open-loop alternatives. Finally, we conclude this section with statistical verification of controllers\footnote{Our results show that integrating a waypoint function in Algorithm~\ref{algo:training} enhances the efficiency of the training process to a small extent.
}.

First, we provide a brief summary of results on evaluation of Algorithm~\ref{algo:training}. Following this, we elaborate on the specifics of our experimental configuration later in this section.

\mypara{Evaluation metric} We evaluate the effectiveness of our methodology outlined in Algorithm~\ref{algo:training} through four case studies, each presenting unique challenges.
First, we present two case studies involving tasks with long time horizons:
\begin{itemize}[wide, labelwidth=!, labelindent=0pt,nosep]
    \item $6$-dimensional quad-rotor combined with a moving platform with task horizon $\horizon = 1500$ time-steps.
    \item $2$-dimensional Dubins car with task horizon $\horizon = 1000$ time-steps.
\end{itemize}
Subsequently, we present two additional case studies characterized by high-dimensional state spaces:
\begin{itemize}[wide, labelwidth=!, labelindent=0pt,nosep]
    \item $20$-dimensional Multi-agent system of $10$ connected Dubins cars with task horizon $\horizon = 60$ time-steps.
    \item $12$-dimensional quad-rotor with task horizon $\horizon = 45$ time-steps. 
\end{itemize}
Table~\ref{tbl:casestudies} highlights the versatility of Algorithm~\ref{algo:training} in handling above case studies. We use a diverse set of temporal tasks which include nested temporal operators and \navid{two} independently moving objects (quad-rotor \& moving platform case study). The detail of the experiments are also discussed as follows.

\begin{table*}
\centering
\resizebox{0.95\textwidth}{!}{$
\begin{tabular}{cccccccc} 
\toprule
Case  Study& Temporal & System   & Time  & NN Controller & Number of  & Runtime   & Optimization Setting\\
          &   Task    & Dimension& Horizon  & Structure     & Iterations & (seconds) & $[M, N, N_1, N_2, \epsilon, b]$      \\
\midrule 
\rowcolor{Gray}
12-D Quad-rotor     & $\varphi_3$ & 12 & 45 steps    & [13,20,20,10,4] & 1120  &  6413.3 & [9,\ 5,\ 30,\ 40,\ $10^{-5}$,\ 5] \\
Multi-agent    & $\varphi_4$ & 20 & 60 steps  & [21,40,20] &  2532 & 6298.2  &[12,\ 5,\ 30,\ 1,\ $10^{-5}$,\  15]\\
\rowcolor{Gray}
6-D Quad-rotor \& Frame & $\varphi_5$ & 7  & 1500 steps  & [8,20,20,10,4] & 84 &  443.45 & [100,\ 15,\ 30,\ 3,\ $10^{-5}$,\ 15] \\
Dubins car & $\varphi_6$ & 2  & 1000 steps  & [3,20,2] & 829 &  3728 & [200,\ 5,\ 60,\ 3,\ $10^{-5}$, 15] \\
\bottomrule
\end{tabular}
$}
\caption{Results on different case studies. \navid{Here, $b$ is the hyper-parameter we utilized to generate $\lbfortl$ in \cite{hashemi2024scaling}. }}\label{tbl:casestudies}
\vspace{-9mm}
\end{table*}
\subsection{12-dimensional Quad-rotor (Nested 3-Future Formula)}

\begin{wrapfigure}[20]{r}{0.45\textwidth}
    \vspace{-2mm}
    \hspace{2mm}
    \includegraphics[trim={1.1cm
  0.2cm 1.1cm 0.2cm},clip,width=0.95\linewidth]{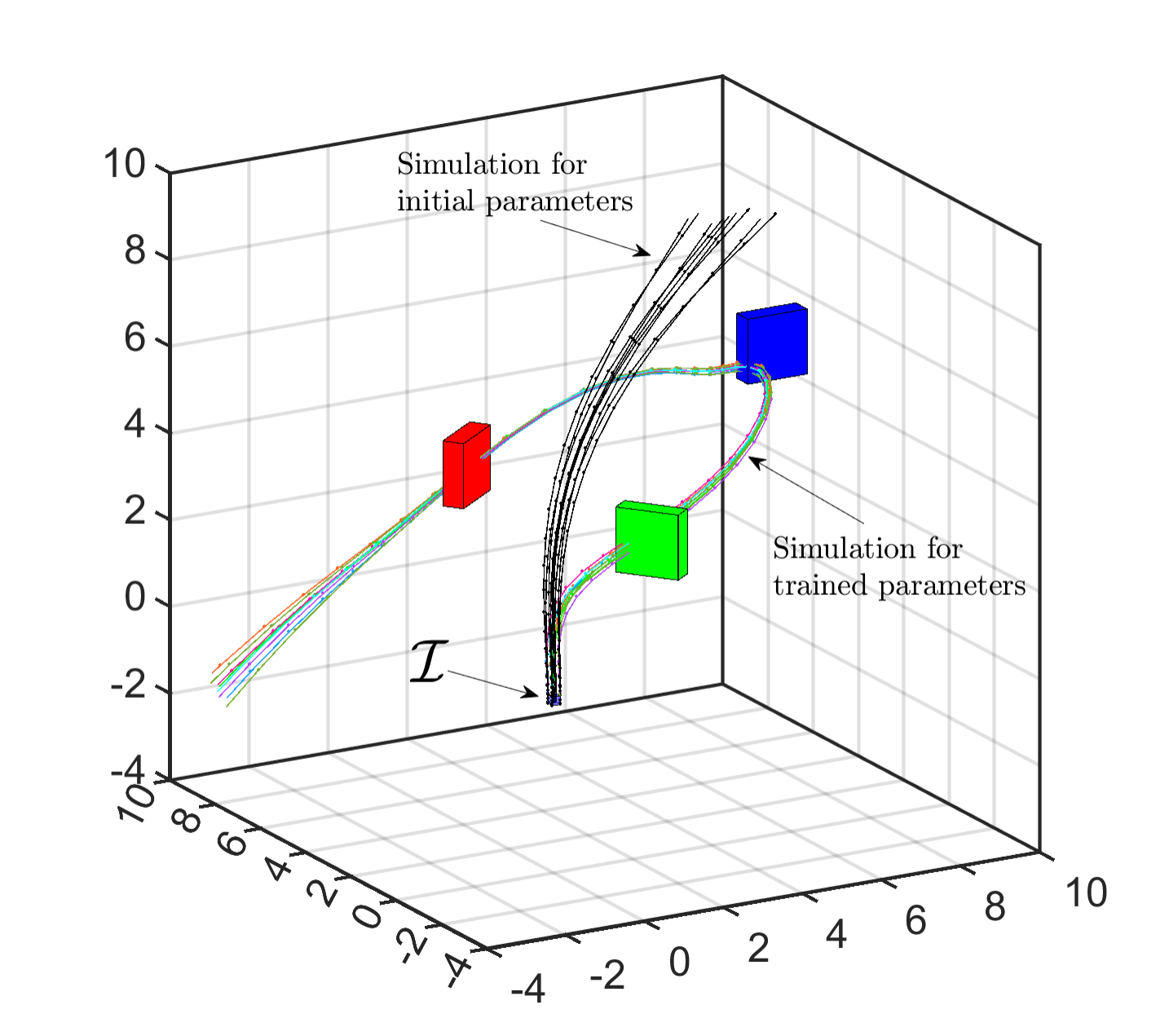}
    \captionsetup{skip=0mm}
    \caption{This figure shows the simulation of trained control parameters to satisfy the specified temporal task in companion with the simulation result for initial guess for control parameters.}
    \label{fig:Quad12intro}
\end{wrapfigure}
We assume a $12$-dimensional model for the quad-rotor of mass, $m=1.4 \ \text{kg}$. The distance of rotors from the quad-rotor's center is also $\ell = 0.3273 \ \text{meter}$ and the inertia of vehicle is $J_x = J_y= 0.054$ and $J_z=0.104$ (see \citep{beard2008quadrotor} for the detail of quad-rotor's dynamics). The controller sends bounded signals $\delta_r,\delta_l,\delta_b, 0\leq \delta_f \leq 1$ to the right, left, back and front rotors respectively to drive the vehicle. Each rotor is designed such that given the control signal $\delta$ it generates the propeller force of $k_1\delta$ and also exerts the yawing torque $k_2 \delta$ into the body of the quad-rotor. We set $k_1= 0.75 mg$ such that, the net force from all the rotors can not exceed $3$ times of its weight, $(g =9.81)$. We also set $k_2 = 1.5 \ell k_1$ to make it certain that the maximum angular velocity in the yaw axis is approximately equivalent to the maximum angular velocity in the pitch and roll axis. We use the sampling time $\delta t = 0.1 \ \mathrm{seconds}$ in our control process. The dynamics for this vehicle is proposed in Eq.~\eqref{eq:Quad12}, where $F, \tau_\phi, \tau_\theta, \tau_\psi$ are the net propeller force, pitch torque, roll torque and yaw torque respectively. We plan to train a NN controller with $\mathbf{tanh}()$ activation function and structure $[13,20,20,10,4]$ for this problem that maps the vector, $[\statee_\timeid^\top , \timeid]^\top$ to the unbounded control inputs $[a_{1,\timeid},a_{2,\timeid}, a_{3,\timeid}, a_{4,\timeid}]^\top$. \navidg{In addition to this, the trained controller should be valid for all initial states,
\[
\init= \left\{ \statee_0 \mid \left[-0.1, -0.1,-0.1, \vec{0}_{9\times1}\right]^\top \leq \statee_0 \leq \ \ \left[\ \ 0.1,\ \  0.1,\ \ 0.1, \vec{0}_{9\times1}\right]^\top \right\}
\vspace{-6mm}
\]}
\begin{figure*}
\begin{equation}\label{eq:Quad12}
\resizebox{\linewidth}{!}{$
{\tiny 
    \hspace{0mm}\resizebox{\hsize}{!}{$\left\{ \hspace{-3mm} \begin{array}{ll}
    &\dot{x}_1 = \cos(x_8) \cos(x_9)x_4+( \sin(x_7) \sin(x_8) \cos(x_9)- \cos(x_7) \sin(x_9))x_5\\
    &\hspace{2mm}+( \cos(x_7) \sin(x_8) \cos(x_9)+ \sin(x_7) \sin(x_9))x_6\\
    &\dot{x}_2 = \cos(x_8) \sin(x_9)x_4+( \sin(x_7)  \sin(x_8)  \sin(x_9)+ \cos(x_7)  \cos(x_9)) x_5\\
    &\hspace{2mm}+( \cos(x_7)  \sin(x_8)  \sin(x_9)- \sin(x_7)  \cos(x_9)) x_6\\
    &\dot{x}_3 =  \sin(x_8) x_4- \sin(x_7)  \cos(x_8) x_5- \cos(x_7)  \cos(x_8) x_6\\
    &\dot{x}_4 = x_{12} x_5-x_{11} x_6-9.81  \sin(x_8)\\
    &\dot{x}_5 = x_{10} x_6-x_{12} x_4+9.81  \cos(x_8)  \sin(x_7)\\
    &\dot{x}_6 = x_{11} x_4-x_{10} x_5+9.81  \cos(x_8)  \cos(x_7)-F/m\\
    &\dot{x}_7 = x_{10}+( \sin(x_7) ( \sin(x_8)/ \cos(x_8))) x_{11}+( \cos(x_7) ( \sin(x_8)/ \cos(x_8))) x_{12}\\
    &\dot{x}_8 =  \cos(x_7) x_{11}- \sin(x_7) x_{12}\\
    &\dot{x}_9 = ( \sin(x_7)/ \cos(x_8)) x_{11}+( \cos(x_7)/ \cos(x_8)) x_{12}\\
    &\dot{x}_{10} = -((J_y-J_z)/J_x)x_{11} x_{12}+(1/J_x) \tau_\phi\\
    &\dot{x}_{11} = ((J_z-J_x)/J_y) x_{10} x_{12}+(1/J_y)) \tau_\theta\\
    &\dot{x}_{12} = (1/J_z) \tau_\psi
    \end{array}
    \right. 
    \begin{array}{l}
    \begin{bmatrix} F \\ \tau_\phi \\ \tau_\theta \\ \tau_\psi \end{bmatrix} = \begin{bmatrix} k_1 & k_1 & k_1 & k_1 \\ 0  & -\ell k_1  & 0 & \ell k_1 \\ \ell k_1 & 0 & -\ell k_1 & 0 \\ -k_2  &  k_2 & -k_2 & k_2 \end{bmatrix} \begin{bmatrix} \delta_f \\ \delta_r \\ \delta_b \\ \delta_l \end{bmatrix}\\\\
    \delta_f = 0.5(\tanh(0.5\  a_1)+1),\\
    \delta_r = 0.5(\tanh(0.5\  a_2)+1),\\
    \delta_b = 0.5(\tanh(0.5\  a_3)+1),\\
    \delta_l = 0.5(\tanh(0.5\  a_4)+1),\\
    a_1, \ a_2,\ a_3,\ a_4 \in \mathbb{R}.
    \end{array}$}}
    $}
\end{equation}
\end{figure*}

Figure~\ref{fig:Quad12intro} shows the simulation of quad-rotor's trajectories with our trained controller parameters. The quad-rotor is planned to pass through the green hoop, \navid{between the $10$th and $15$th time-step}. Once it passed the green hoop it should pass the blue hoop in the future $10$th to $15$th time-steps and again once it has passed the blue hoop it should pass the red hoop again in the future next $10$ to $15$ time-steps. This is called a nested future formula, in which we design the controller such that the quad-rotor satisfies this specification. Assuming $p$ as the position of quad-rotor, this temporal task can be formalized in DT-STL framework as follows:
\begin{equation}
\varphi_3 = \ev_{[10,15] } \left(\  p \in \text{green\_hoop} \  \wedge  
            \ev_{[10,15] } \left(\ p \in \text{blue\_hoop}\ \wedge 
            \ev_{[10,15] } \left(\  p \in \text{red\_hoop} \ \right)\ 
            \right)\ 
   \right)\ 
\end{equation}

Figure~\ref{fig:Quad12intro} shows the simulation of trajectories, generated by the trained controller. The black trajectories are also the simulation of the initial guess for the controller, which are generated completely at random and are violating the specification. We sampled $\init$ with $9$ points, that are the corners of $\init$ including its center. The setting for gradient sampling is $M=9,\ N=5$. We trained the controller with $\bar{\rob} =0$, in Algorithm~\ref{algo:training} with optimization setting $(N_1=30,\ N_2=40,\ \epsilon=10^{-5})$ over $1120$ gradient steps (runtime of $6413.3\ \mathrm{seconds}$). The runtime to generate $\lbfortl$ is also $0.495 \ \mathrm{seconds}$ and we set $b=5$ \navid{for it}. The Algorithm~\ref{algo:training}, utilizes gradients from waypoint function, critical predicate, and $\lbfortl$ , $515, 544$, and $61$ times respectively.

\subsection{Multi-Agent: Network of Dubins Cars (Nested Formula)}\label{sec:multiagent}

\begin{figure*}[t] \begin{subfigure}[b]{0.34\textwidth}
    \begin{subfigure}{\linewidth} \centering \includegraphics[width
    =0.95\linewidth]{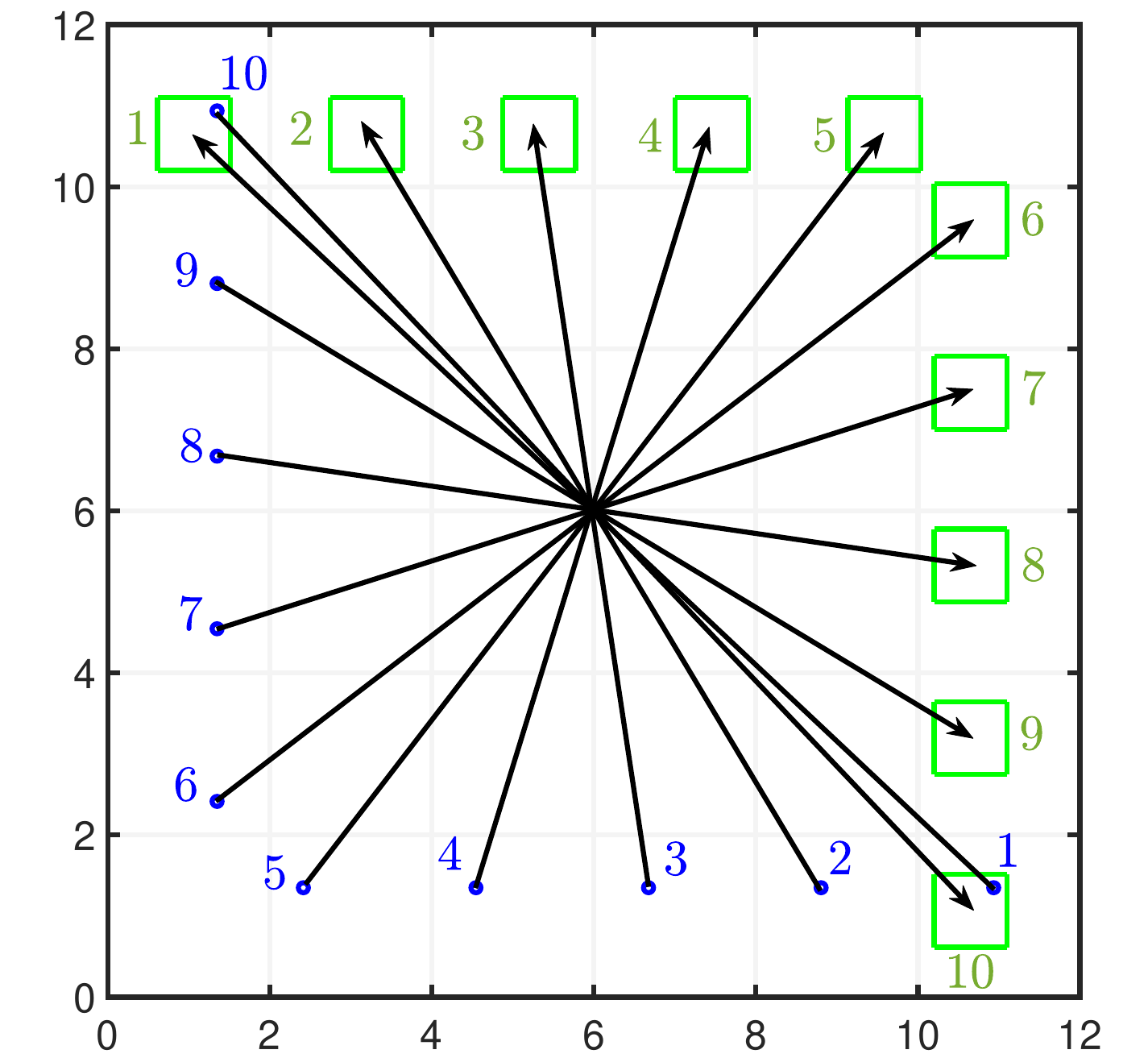} \captionsetup{skip=0mm} \caption{agents
    vs goal sets} \label{fig:multiplan} \end{subfigure} \vfill \vspace{2mm}
    \begin{subfigure}{\linewidth} \centering
      \includegraphics[width=0.95\linewidth]{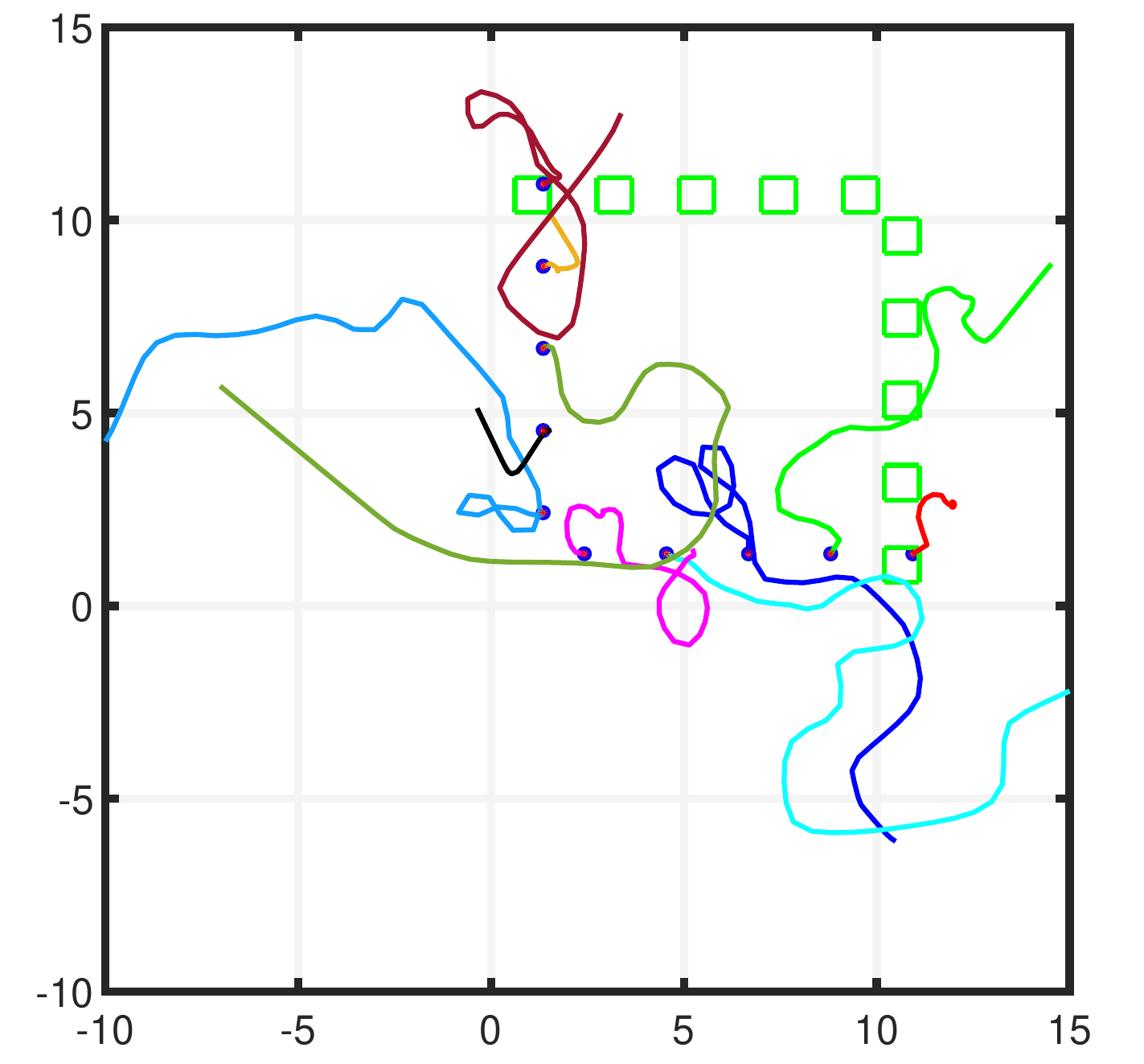}
      \captionsetup{skip=0mm} \caption{initial guess for $\param^{(0)}$}
      \label{fig:initplan} \end{subfigure} \end{subfigure} \hfill
  \begin{subfigure}[b]{0.64\textwidth} \centering \includegraphics[trim={1.1cm
  0.2cm 1.45cm 0.2cm},clip,width=\textwidth]{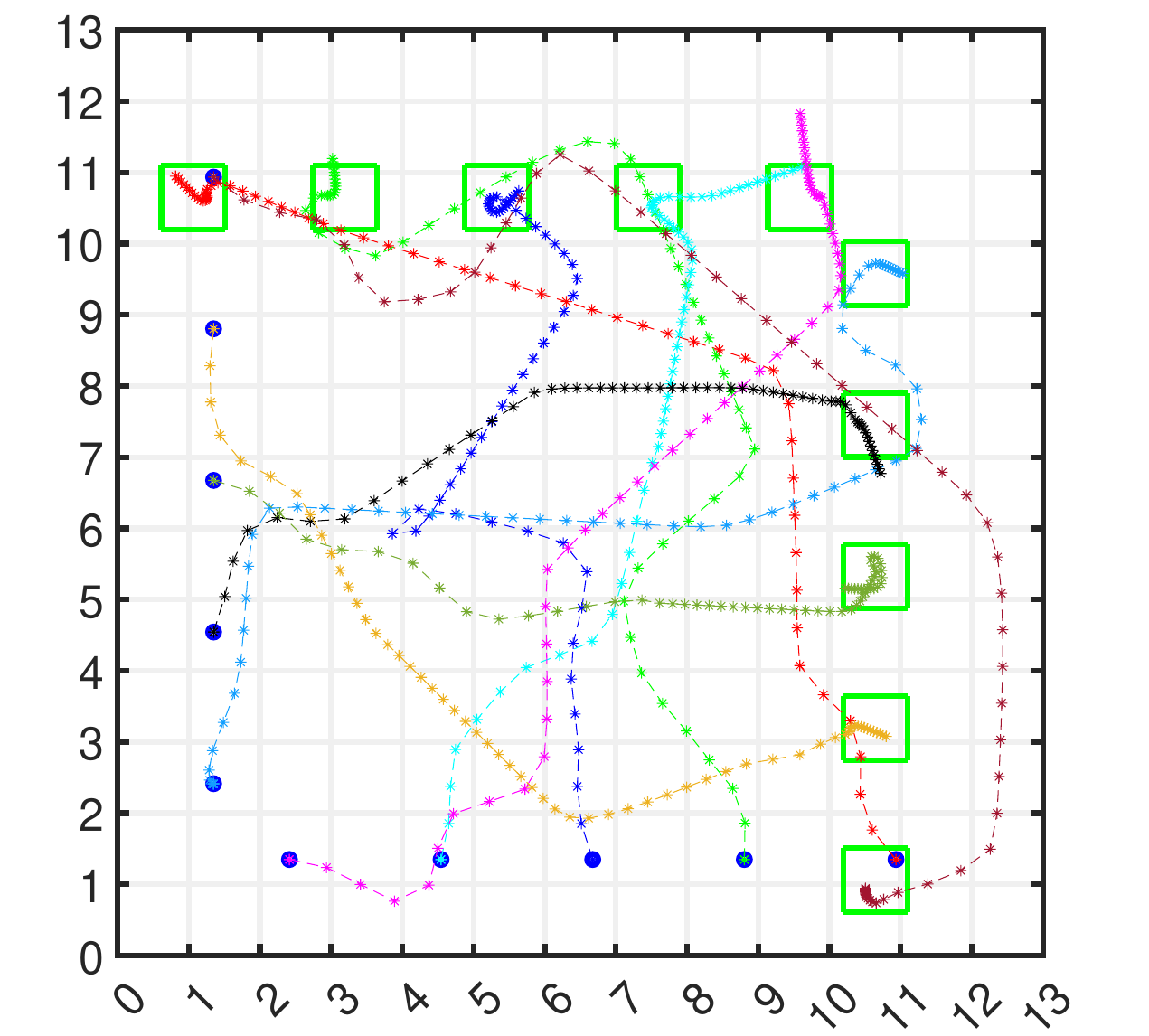}
    \captionsetup{skip=0mm} \caption{Simulation of trajectories for trained
    control parameters.} \label{fig:trainedplan} \end{subfigure}
\captionsetup{skip=1mm}
\caption{\small These figures show a multi-agent system of 10 connected Dubins
cars. Fig.~(a) shows the start (blue dots) and goal points (green squares) for
agents. Figs.~(b,c) show simulated system trajectories with both the initial
untrained controller and the centralized NN controller trained with
Algorithm~\ref{algo:training}. The controller coordinates all cars to reach
their respective goals between $20$ and $48$ seconds, and then stay in their
goal location for at least $12$ seconds. It also keeps the cars at a minimum
distance from each other. We remark that the agents finish their tasks (the
first component of $\varphi_4$) at different times.}
\end{figure*}

\newcommand{\vx}{\mathbf{x}}
In this example, we assume a network of $10$ different Dubins cars that are all under the control of a neural network controller. The dynamics of this multi-agent system is,
\begin{equation}
\label{eq:dubins}
    \begin{bmatrix}\dot{x}^i\\ \dot{y}^i \end{bmatrix}=\begin{bmatrix} v^i \cos(\theta^i)\\
    v^i \sin(\theta^i)\end{bmatrix}, \qquad   
    \begin{aligned}
     &v^i \gets\    \tanh(0.5 a_1^i)+1, a_1^i \in \mathbb{R}\\
     &\theta^i \gets  a_2^i \in \mathbb{R}
    \end{aligned},\qquad \ i\in 1,\cdots,10,
\end{equation}
that is, a 20 dimensional multi-agent system with $20$ controllers, $0 \leq v^i \leq 1 \theta^i\in \mathbb{R},\  i\in 1,\cdots,10$. Figure~\ref{fig:multiplan} shows the initial position of each Dubins car in $\mathbb{R}^2$ in companion with their corresponding goal sets. The cars should be driven to their goal sets, and they should also keep a minimum distance of $d=0.5$ meters from each other while they are moving toward their goal sets. We assume a sampling time of $\delta t = 0.26\  \mathrm{seconds}$ for this model, and we plan to train a NN controller with $\mathbf{tanh}()$ activation function and structure $[21,40,20]$ via Algorithm~\ref{algo:training}. \navidg{ For this problem, the controller maps the vector}, $[\statee_\timeid^\top, \timeid]^\top$ to the unbounded control inputs $\left\{a_{1,\timeid}^i,a_{2,\timeid}^i\right\}_{i=1}^{10}$.  Note that $\statee^i_\timeid = (x^i_\timeid,y^i_\timeid)$.
This temporal task can be formalized in DT-STL framework as follows:
\[
\varphi_4 :=  
    \bigwedge_{i=1}^{10}  
        \left(\ev_{[20,48] } 
            \alw_{[0,12] } \left( \statee^i \in \text{Goal}^i \right)\right) 
    \qquad\bigwedge\qquad
        \bigwedge_{\underset{i,j \in  \left\{1,\cdots,10\right\}}{i \neq j}}
            \alw_{[0,60] } 
                \left(\| \statee^i - \statee^j \|_\infty > d \right)
\]
Figure~\ref{fig:trainedplan} shows the simulation of the trajectories for the trained controller, and Figure~\ref{fig:initplan} presents the simulation of trajectories for the initial guess for control parameters. We observe that our controller manages the agents to finish the task in different times. \navidg{Thus, we present the time-stamps with asterisk markers} to enhance the clarity of the presentation regarding satisfaction of the specification in Figure~\ref{fig:trainedplan}. Although the task is not a long horizon task, due to the high dimension and complexity of the task, we were unable to solve this problem without time sampling. However, we successfully solved this problem with Algorithm~\ref{algo:training} within $6298\ \mathrm{seconds}$ and $2532$ gradient steps.

We also set the optimization setting as, $M=12, N=5, N_1 = 30, N_2=1, \epsilon = 10^{-5}$. The runtime to generate $\lbfortl$ is also $6.2\ \mathrm{seconds}$ \navid{and we set $b=15$ for it}. Over the course of the training process we utilized $187, \ 1647$ and $698$ gradients from waypoint function, critical predicate and $\lbfortl$ respectively.

\subsection{6-dimensional Quad-rotor \& Moving Platform: Landing a Quad-rotor}\label{sec:dronemission}

We use a $6$-dimensional model for quad-rotor dynamics as follows.
\begin{equation}
\label{eq:QR}
\begin{aligned}
    &\begin{bmatrix}\dot{x}& \dot{y}& \dot{z}& \dot{v}_x& \dot{v}_y& \dot{v}_z\end{bmatrix}=\begin{bmatrix}v_x&v_y&v_z& \mathrm{g} \tan(u_1)&-\mathrm{g} \tan(u_2) & \mathrm{g}-u_3  \end{bmatrix},\ \text{where,} \\
    &u_1 \gets\   0.1 \tanh(0.1 a_1),\quad    u_2 \gets\   0.1 \tanh(0.1 a_2),\quad u_3 \gets\   \mathrm{g}-2 \tanh(0.1 a_3),\quad  a_1,a_2,a_3\in \mathbb{R}.
\end{aligned}
\end{equation}  
\navidg{Let $\mathbf{x} = (x,y,z)$ denote the quad-rotor's position 
and $\mathbf{v} = (v_x,v_y,v_z)$ denote its velocity
along} the three coordinate axes. The control inputs $u_1, u_2, u_3$ represent the pitch, roll, and thrust inputs respectively. We assume that the inputs are bounded as follows: $-0.1 \leq u_1,u_2 \leq 0.1, \ 7.81 \leq u_3 \leq 11.81$. 

\begin{figure*}
\begin{minipage}{0.6\textwidth}
    \includegraphics[width=0.9\linewidth]{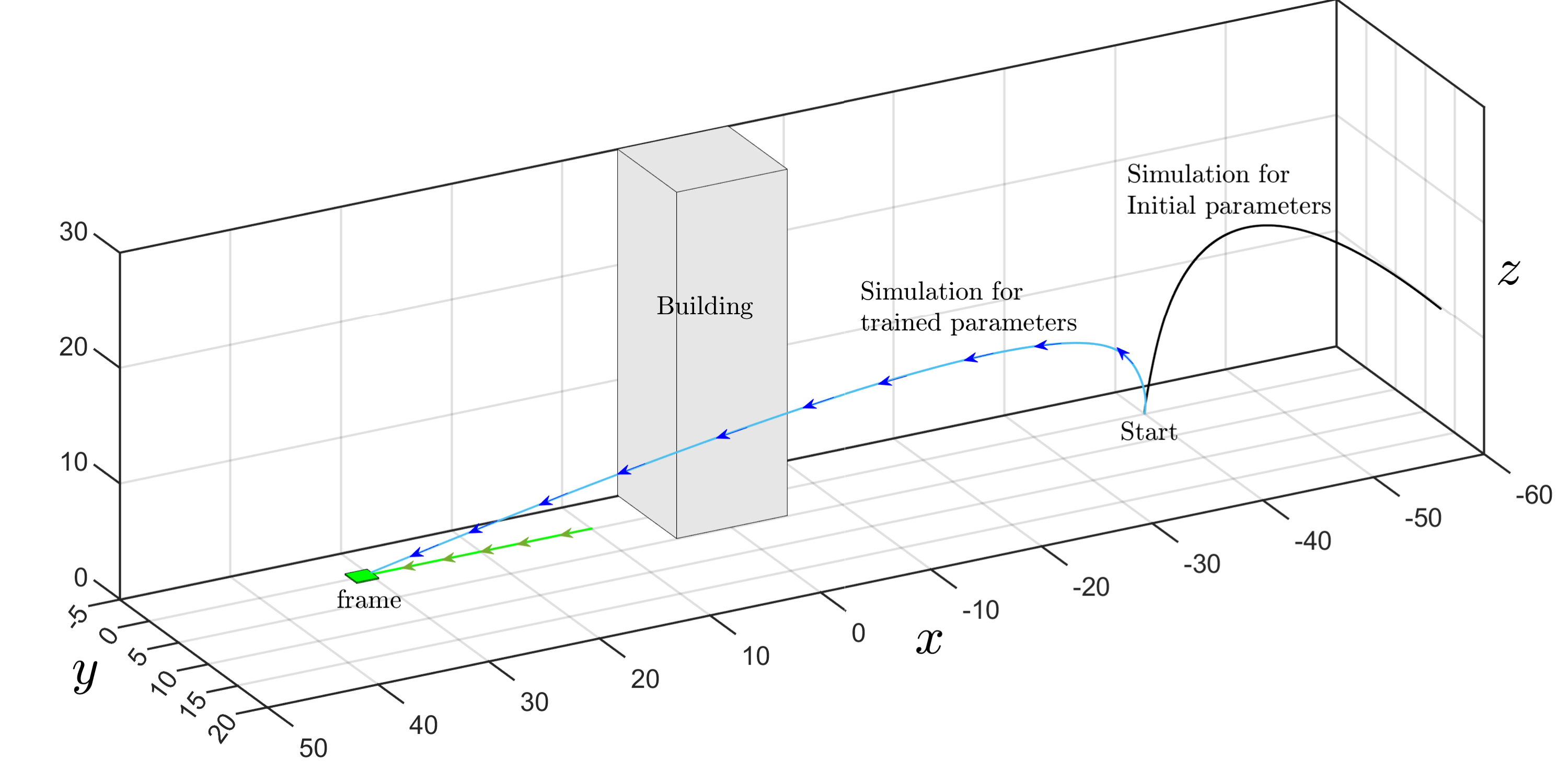}
    \captionsetup{skip = -1mm}
    \caption{This figure shows the simulated trajectory for trained controller in comparison to the trajectories for naive initial random guess. The frame is moving with a velocity determined with the controller that also controls the quad-rotor.}
    \label{fig:dronemission}
\end{minipage}
  \hfill
\begin{minipage}{0.36\textwidth}
    \begin{equation}\label{eq:goalset}
    \hspace{-14mm}
    \resizebox{\linewidth}{!}{$
    \text{Goal} =\left\{ \begin{bmatrix} x_\timeid \\ y_\timeid\\ z_\timeid \\ v_{x,\timeid} \\ v_{y,\timeid}\\ v_{z,\timeid} \\ x^f_{\timeid} \end{bmatrix} \mid  \begin{bmatrix} -1 \\ -1\\ 0.11 \\ 0 \\ -1\\ -1 
    \end{bmatrix}     \leq     
    \begin{bmatrix} x_\timeid-x^f_\timeid \\ y_\timeid\\ z_\timeid \\ v_{x,\timeid} \\ v_{y,\timeid}\\ v_{z,\timeid} 
    \end{bmatrix}      \leq      \begin{bmatrix}  1 \\  1\\ 0.6 \\  2 \\  1\\  1 
    \end{bmatrix} \right\} $}
    \end{equation}
\end{minipage}
\end{figure*}

The horizon of the temporal task is $1500$ time-steps  with $\delta t = 0.05s$. The
quad-rotor launches at a helipad located at $(x_0,y_0,z_0)=(-40 , 0, 0)$. We
accept a deviation of $0.1$ for $(x_0$ and $y_0$ and train the controller to
be valid for all the states sampled from this region. The helipad is also $40m$
far from a building located at $(0,0,0)$. The building is $30m$ high, where the
building's footprint is $10m \times 10m$. We have also a moving platform with
dimension $2m\times 2m\times 0.1m$ that is starting to move from $(10,0,0)$ with
a variable velocity, modeled as, $\ \dot{x}^f =  u_4$. We accept a deviation of
$0.1$ for $x^f_0$, and our trained controller is robust with respect to this
deviation. We define $\sampledinits$ with $9$ samples located at the corners of
$\init$ and the center of $\init$. The frame is required to keep a minimum
distance of $4.5$ meters from the building. We train the NN controller to
control both the quad-rotor and the platform to ensure that the quad-rotor will
land on the platform with relative velocity of at most $1\ m/s$ in $x,y$ and $z$
directions, and its relative distance is at most $1m$ in $x,y$ direction and
$0.4m$ in $z$ direction. Let $p = (x,y,z)$ be the position of the quad-rotor,
this temporal task can be formulated as a reach-avoid formula in DT-STL
framework as follows:
\begin{equation}
\varphi_5 =  \alw_{[0,1500] } \left( p \notin \text{obstacle}\right)\ \wedge \ev_{[1100,1500] } ( p \in \text{Goal})\ \wedge \alw_{[0,1500] }(x^f_\timeid>9.5)
\end{equation}
where the goal set is introduced in \eqref{eq:goalset}.
We plot the simulated trajectory for the center of set of initial states $\init$, in Figure~\ref{fig:dronemission}. The NN controller's structure is specified as $[8, 20,20,10,4]$ and \navidg{uses $\tanh()$  activation function}. We initialize it with a random guess for its parameters. The simulated trajectory for initial guess of parameters is also depicted in black. The setting for gradient sampling is $M=100,\ N=15$. We trained the controller with $\bar{\rob} =0$, over $84$ gradient steps (runtime of $443\ \mathrm{seconds}$). The runtime to generate $\lbfortl$ is also $7.74 \ \mathrm{seconds}$ and we set $b=15$ \navid{, for it}. In total, the Algorithm~\ref{algo:training}, utilizes gradients from waypoint function, critical predicate, and $\lbfortl$ , $5, 71$, and $8$ times respectively.
\subsubsection{Influence of waypoint function, critical predicate and time sampling on Algorithm~\ref{algo:training} }\label{apdx:analysis}
\begin{figure}[t]
    \includegraphics[width=1.01\linewidth]{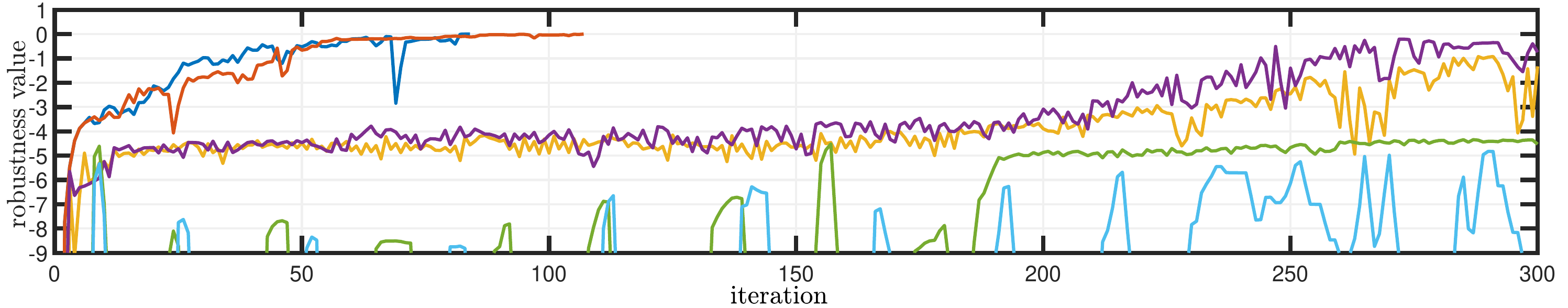}
    \captionsetup{skip=0mm}\caption{This figure shows the learning curve for training processes. Note, the figure has been truncated and the initial robustness for all the experiments at iteration $0$ is $-47.8$. This figure shows that Algorithm~\ref{algo:training} in the presence of the waypoint function concludes successfully in 84 iterations while when the waypoint function is not included, it terminates in $107$ iterations. The algorithm also fails if the critical predicate is not considered in time sampling.}\label{fig:comparison}
\end{figure}

Here, we consider the case study of landing a quad-rotor, and perform an ablation study over the impact of including 1) critical predicate, 2) waypoint function and 3) time sampling, in the training process via Algorithm~\ref{algo:training}. To that end, we compare the results once these modules are excluded from the algorithm. In the first step, we remove the waypoint function and show the performance of the algorithm. In the next step, we also disregard the presence of critical time in time-sampling and train the controller with completely at random time-sampling, and finally we examine the impact of time sampling on the mentioned results. Table~\ref{tbl:comparison} shows the efficiency of training process in each case, and Figure~\ref{fig:comparison} compare the learning curves. Our experimental result shows, the control synthesis for quad-rotor (landing mission) faces a small reduction in efficiency when the waypoint function is disregarded and fails when the critical predicate is also removed from time sampling. This also shows that control synthesis fails when time-sampling is removed.
\begin{table*}
\centering
\begin{tabular}{cccccc}
    \toprule 
Learning curve's                & Waypoint &   Critical & Time- & Number of & Runtime \\
color in Figure~\ref{fig:comparison}   &function & predicate & sampling & Iterations \\
\midrule \rowcolor{Gray} 
\textcolor{colorone}{\rule{5mm}{1.0 pt}}  &$\checkmark$ & $\checkmark$ & $\checkmark$ &   $84$        & $443 \sec.$ \\
\textcolor{colortwo}{\rule{5mm}{1.0 pt}}  &$\times$     & $\checkmark$ & $\checkmark$ &  $107$        & $607 \sec.$ \\ 
\rowcolor{Gray}
\textcolor{colorthree}{\rule{5mm}{1.0 pt}}&$\checkmark$ & $\times$     & $\checkmark$ &  DNF$[-0.74]$ & $6971 \sec.$ \\
\textcolor{colorfour}{\rule{5mm}{1.0 pt}} &$\times$     & $\times$     & $\checkmark$ &  DNF$[-1.32]$ & $4822 \sec.$ \\
\rowcolor{Gray} 
\textcolor{colorfive}{\rule{5mm}{1.0 pt}} &$\checkmark$ & $\checkmark$ & $\times$     &  DNF$[-4.52]$ & $1505\sec.$  \\ 
\textcolor{colorsix}{\rule{5mm}{1.0 pt}}  &$\times$     & $\checkmark$ & $\times$     &  DNF$[-11.89]$& $1308 \sec.$ \\ 
\bottomrule 
\end{tabular}
\caption{Ablation studies for picking different options for the optimization process. This table shows the results of the training algorithm in case study \ref{apdx:analysis}. We indicate that the training does not result in positive robustness within $300$ gradient steps by DNF (\em{did not finish}) with the value of robustness in iteration $300$ in brackets. The table represents an ablation study, where we disable the various heuristic optimizations in Algorithm~\ref{algo:training} in different combinations and report the extent of reduction in efficiency. We use $\checkmark,\times$ to respectively indicate a heuristic being included or excluded. The time-sampling technique is utilized in
all the experiments.}\label{tbl:comparison}
\end{table*}

\subsection{Dubins Car: Growing Task Horizon for Dubins Car (Ablation study on time sampling) }\label{apdx:experiment}

In this experiment, we consider Dubins car with dynamics,
$$
    \begin{bmatrix}\dot{x}\\ \dot{y} \end{bmatrix}=\begin{bmatrix} v \cos(\theta)\\
    v \sin(\theta)\end{bmatrix}, \quad \ v \gets\    \tanh(0.5 a_1)+1, a_1 \in \mathbb{R}, \ \quad \theta \gets  a_2 \in \mathbb{R},
$$
and present an ablation study on the influence of gradient sampling on control synthesis. Given a scale factor $a >0 $, a time horizon $\horizon$ and a pre-defined initial guess for control parameters $\param^{(0)}$, we plan to train a $\tanh()$ neural network controller with structure $[3,20,2]$, to drive a Dubins car, 
to satisfy the temporal task,\  $\varphi_6:= \ev_{[0.9\horizon, \horizon] } \left( p \in \text{Goal}\right) \wedge \alw_{[0, \horizon] } \left( p \notin \text{Obstacle} \right)$, where $p = (x,y)$ is the position of Dubins car. The Dubins car starts from $(x_0, y_0) = (0,0)$. The obstacle is also a square centered on $(a/2, a/2)$ with the side length $2a/5$. The goal region is again a square centered on $(9a/10, 9a/10)$ with the side length $a/20$. We solve this problem for $\horizon = 10, 50, 100, 500, 1000$ and we also utilize $a = \horizon/10$ for each case study. We apply standard gradient ascent (see Algorithm~\ref{algo:vanilla}) to solve each case study, both with and without gradient sampling. 

\begin{wrapfigure}[17]{r}{0.45\textwidth}
    \vspace{-7mm}
    \hspace{2mm}
    \includegraphics[trim={1.1cm
  0.2cm 1.1cm 0.2cm},clip,width=0.85\linewidth]{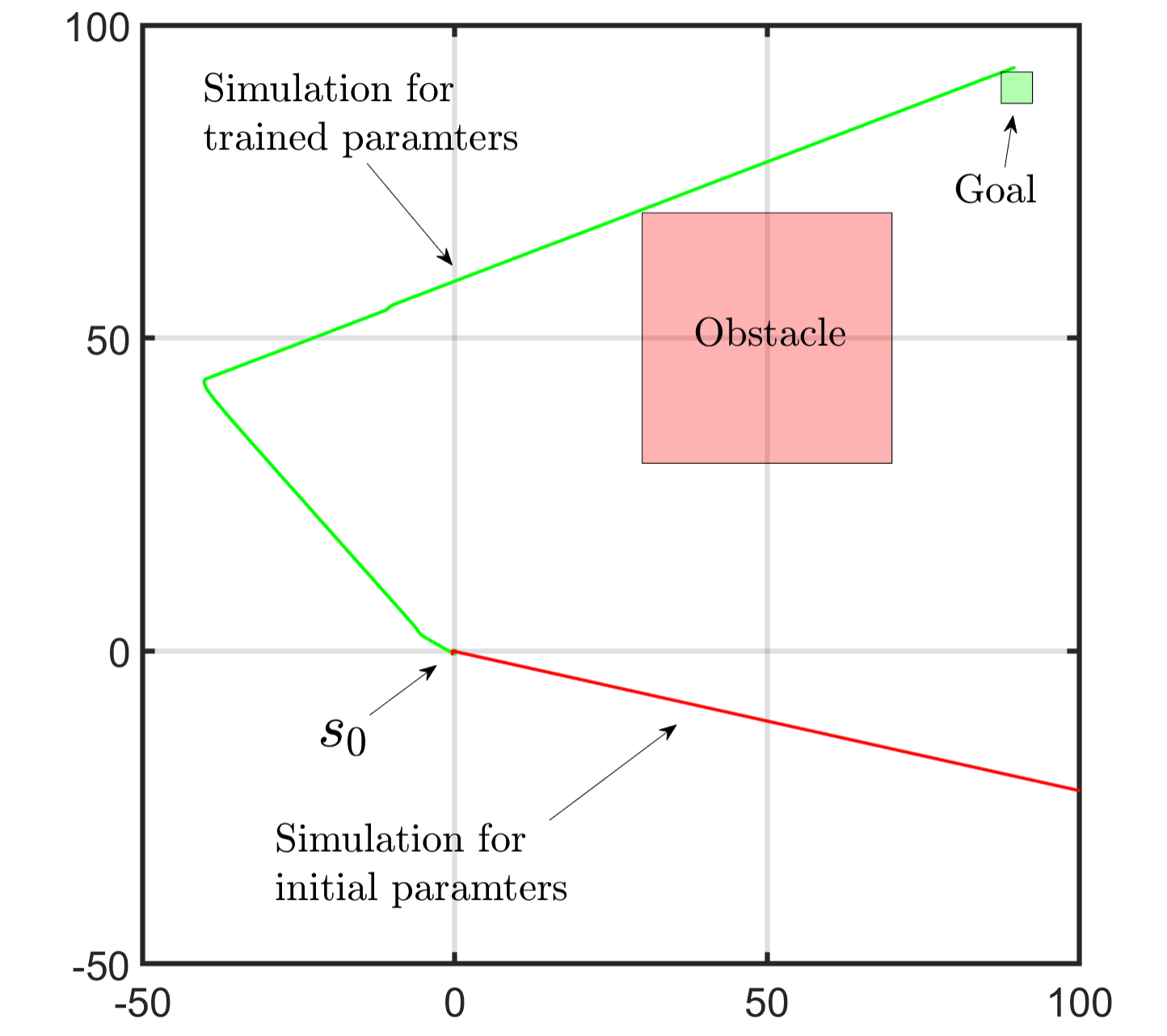}
    \captionsetup{skip=0mm}
    \caption{This figure shows the simulation of the  results for Dubins car in the ablation study proposed in section (\ref{apdx:experiment}). In this experiment, the task horizon is $1000$ time-steps. }
    \label{fig:endresult}
\end{wrapfigure}
Furthermore, in addition to standard gradient ascent, we also utilize Algorithm \ref{algo:training} to solve them. Consider we set the initial guess and the controller's structure similar, for all the training processes, and we also manually stop the process once the number of iterations exceeds 8000 gradient steps. We also assume a singleton as the set of initial states \navidg{$\left\{(0,0)\right\}$} to present a clearer comparison. The runtime and the number of iterations for each training process is presented in Table~\ref{tbl:ablation}. Figure \ref{fig:endresult} displays the simulation of trajectories trained using Algorithm \ref{algo:training} for $K = 1000$ time-steps (via gradient sampling), alongside the  simulations for the initial guess of controller parameters. 

\navid{Table. \ref{tbl:ablation} shows our approximation technique outperforms the original gradient when the computation for original gradient faces numerical issues (such as longer time-horizons $K=500,\ 1000$). However, in case the computation for original gradient does not face any numerical issues, then the original gradient outperforms the sampled gradient which is expected. This table also shows that the standard gradient ascent (with time sampling) is still unable to solve for the case $K=1000$ while the Algorithm \ref{algo:training} solves for this case efficiently. This implies the combination of time-sampling, critical predicate, and safe-resmoothing provides significant improvement in terms of scalability. The experiment $K=500$ in this table also shows, inclusion of waypoint in Algorithm \ref{algo:training} is sometimes noticeably helpful.} 

\begin{table}[t]
\resizebox{0.98\textwidth}{!}{$
\begin{tabular*}{1.4\textwidth}{@{\extracolsep{\fill}}ccccccccc}
\toprule
    &\muc{2}{\navidg{Standard gradient ascent}} &  \muc{2}{\navidg{Standard gradient ascent}} & \muc{2}{Algorithm \ref{algo:training}}\!\!\!\footnotesize{(no waypoint)}  & \muc{2}{Algorithm \ref{algo:training}}\!\!\!\footnotesize{(with waypoint)} \\
Horizon& \muc{2}{\footnotesize{(No time Sampling)}} & \muc{2}{\footnotesize{(With time Sampling)}} & \muc{2}{\footnotesize{(With time Sampling)}} & \muc{2}{\footnotesize{(With time Sampling)}}\\
\cmidrule{2-9}
  &Num. of & Runtime & Num. & Runtime & Num. of & Runtime & Num. of & Runtime \\
 & Iterations & (seconds) &   Iterations & (seconds) &  Iterations & (seconds) &  Iterations & (seconds) \\
\midrule
\rowcolor{Gray} 
10   &  34 & 2.39 &11 & 1.39 &6    &0.9152 & 4  & 5.61   \\
\midrule
50   &  73 & 2.46 &53 &14.01 &20   &2.7063 & 25 & 6.09   \\
\midrule
\rowcolor{Gray} 
100  &  152& 8.65 &105&112.6 &204  &79.33 & 157  & 90.55   \\
\midrule
500  &  DNF$[-1.59]$& 4986   &3237 &8566 & 2569  & 2674   & 624 &890.24 \\
\midrule
\rowcolor{Gray} 
1000 &  DNF$[-11.49]$& 8008  & DNF$[-88.42]$& 28825 & 812  & 1804   & 829& 3728\\
\bottomrule
\end{tabular*}$}\vspace{1mm}\\
\caption{Ablation study. We mark the experiment with DNF[.] if it is unable to provide a positive robustness within $8000$ iterations, and the value inside brackets is the maximum value of robustness it finds. We magnify the environment proportional to the horizon. All experiments for $K=10,50,100$ use a unique guess for initial parameter values, and all the experiments for $K=500,1000$ use another unique initial guess. \navidg{ Here, we utilized critical predicate module in both cases of Algorithm \ref{algo:training} (columns 3 \& 4)}. }
\vspace{-4mm}
\label{tbl:ablation}
\end{table}

\subsection{Robustness of NN feedback controllers over open-loop alternatives}\label{sec:feedback}
\begin{figure}[t]
    \centering
    \includegraphics[width=0.9\linewidth]{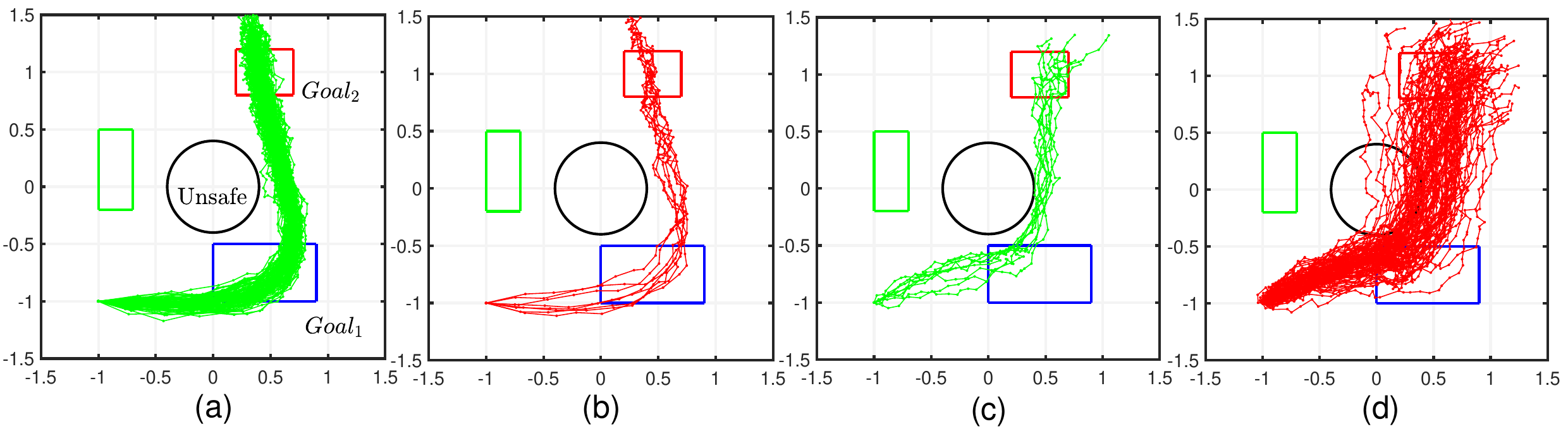}
    \caption{This figure shows the simulation of trajectories when the trained controller is deployed on the noisy deployment environment, both controllers are trained in the presence of noise. The trajectories of NN feedback controller that satisfy (a) and violate (b) the specification and those of the open-loop controller that satisfy (c) and violate (d) the specification are shown.}
    \label{fig:successrate}
\end{figure}

In this section, we empirically demonstrate that feedback NN controllers are
more robust to noise and uncertainties compared to open loop controllers, even
when the feedback controller is not trained in the presence of noise. We then
show that if we train the feedback controller after introducing a stochastic
noise in the original system dynamics, the performance vastly outperforms
open loop control trained in the presence of noise.
To illustrate, we use the example proposed in \cite{leung2019backpropagation} but add a stochastic noise and also include some uncertainty on the choice of initial condition. 
The modified system dynamics are shown in Eq.~\eqref{eq:pert_dyn}, \navidg{where the sampling time $\mathbf{d}t = 0.1$.}
\begin{equation}
\label{eq:pert_dyn}
\statee_{\timeid+1} = \statee_{\timeid} + u_\timeid \mathbf{d}t + c_1 v_\timeid,\qquad \statee_0 = [-1,\ -1]^\top + c_2 \eta.
\end{equation}
\noindent Here, for $k \in 1,\cdots,\horizon$ and $v_k$ and $\eta$ are both i.i.d.
random variables with \navidg{ standard distribution, e.g., $\eta, v_\timeid \sim \mathcal{N}(0_{2\times1}, I_{2\times2})$, where $I_{2\times2}$ is the identity matrix}. In this example, the desired objective for the system is: 
\begin{equation}
\varphi_8 = \ev_{[0,44] }\left( \alw_{[0,5] } \left(\statee \in \text{Goal}_1 \right) \right) \bigwedge \ev_{[0,44] }\left(\alw_{[0,5] } \left(\statee\in\text{Goal}_2 \right) \right) \bigwedge \alw_{[0,49] }  \neg \left( \statee \in \text{Unsafe} \right),
\end{equation}
where the regions $\text{Goal}_1, \text{Goal}_2$, and $\text{Unsafe} $ are illustrated in Figure~\ref{fig:successrate} \footnote{We also add the following updates to the original problem presented in \cite{leung2019backpropagation}:
\begin{itemize}[nosep,leftmargin=1.2em]
    \item We omit the requirement $\statee_{\horizon} = [1,\ 1]^\top$ from both 
    control problems for simplicity.
    \item We increase the saturation bound of the controller to $u_\timeid \leq 4\sqrt{2}$. We also apply this condition to the open-loop controller proposed in \cite{leung2019backpropagation}.
\end{itemize}}.

In the first step of the experiment, we train the feedback and
open-loop controllers in the absence of the noise ($c_1=c_2=0$) and deploy the
controllers on the noisy environment ($c_1 = 0.0314, c_2 = 0.0005$) and compare
their success rate\footnote{To report the success rate, we deploy the
controllers $1000$ different times and compute the percentage of the
trajectories that satisfy the specification.}.  In the second step of the
experiment, we train both the feedback and open-loop controllers on the noisy
environment  ($c_1 = 0.0314, c_2 = 0.0005$), and also deploy them in the
noisy environment ($c_1 = 0.0314, c_2 = 0.0005$) to compare their success rate.
If we train the open-loop and NN feedback controllers in the absence of noise, then the controllers will respectively satisfy the specification in $3.7\%$ and $65.4\%$ of trials when deployed in a noisy environment. However, we can substantially improve performance of feedback controllers by training in the presence of noise; here, the controllers satisfy the spec $5.4\%$ and $94.4\%$ of trials respectively showing that the NN feedback controller has better overall
performance in the presence of noise, which open-loop control lacks.

\begin{wrapfigure}[15]{r}{.55\textwidth}
    \vspace{-4mm}
    \begin{mdframed}
    \begin{algorithm}[H]
    \small{
    \textbf{ Initialize variables}\;
    \While{$\left(\underset{\statee_0 \in \sampledinits}{\min} 
              \left( \rob(\varphi,\traj[\statee_0\ ;{\param^{(j)}}],0) \right) < \bar{\rob}\right)$}{
         $\statee_0 \gets$ $\mathtt{Sample\ from\ } \sampledinits$\;
             $\traj[\statee_0\ ;{\param^{(j)}}] \gets$ $\mathtt{Simulate\ using\ policy\ } \policy_{\param^{(j)}}$\;
             $d \gets \gradient{\stltolb(\traj[\statee_0\ {\param^{(j)}}])}{\param} \ \mathtt{using\ } \traj[\statee_0\ ;{\param^{(j)}}]$\; 
             $\param^{(j+1)} \gets \param^{(j)}+\adam(d)$ \;
             $j \gets j+1$\;
    }}
    \caption{\navidg{Standard} Gradient Ascent Backpropagation via smooth semantics}\label{algo:vanilla}
    \end{algorithm}
    \end{mdframed}
\end{wrapfigure}
We utilized STLCG PyTorch toolbox \cite{leung2019backpropagation} to solve for
the open-loop controller. We also utilized the standard gradient ascent proposed
in Algorithm~\ref{algo:vanilla} (via $\lbfortl$ as smooth semantics $\stltolb$)
for training the feedback controllers. We let the training process in
Algorithm~\ref{algo:vanilla} and STLCG to run for $5000$ iterations, and then
terminated the training process.  Figure~\ref{fig:successrate} shows the
simulation of trained controllers when they are deployed to the noisy
environment. Here, we generate $100$ random trajectories via trained
controllers and plot them in green and red when they satisfy or violate the
specification, respectively. However, all trained feedback controllers in this paper exhibit the same level of robustness to noise. 

\subsection{Statistical verification of synthesized controllers}\label{sec:verific}

In \cite{hashemi2023neurosymbolic}, we showed that if the trained neural network controller, the plant dynamics and the neural network representing the STL quantitative semantics all use $\relu$ activation functions, then we can use tools such as NNV \cite{tran2020nnv} that compute
the forward image of a polyhedral input set through a neural network to verify whether a given DT-STL property holds {\em for all} initial states of the system. However, there are few challenges in applying such deterministic methods here: we use more general activation functions, the depth of the overall neural network can be significant for long-horizon tasks, and
the dimensionality of the state-space can also become a bottleneck. In this paper, we thus eschew the use of deterministic techniques, instead reasoning about the correctness of our neural network feedback control scheme using a statistical verification approach. In other words, given the coverage level $\delta_1 \in (0,1)$ and confidence level $\delta_2 \in (0,1)$ we are interested in a probabilistic guarantee of the form, $ \Pr[\ \ \ \Pr[\traj[\statee_0\ ;\param] \models \varphi]\ \ \  \geq \delta_1\ \ \ ]\ \ \  \geq \delta_2$.

The main inspiration for our verification is drawn from the theoretical developments in conformal prediction \cite{vovk2005algorithmic}. Of particular significance to us is the following lemma:
\newcommand{\BetaD}{\mathbf{Beta}}
\begin{lemma}[From \cite{david2004order}]
\label{lem:vovk}
Consider $m$ independent and identically distributed (i.i.d.), real-valued data points drawn from some distribution $\mathcal{D}$. After they are drawn, suppose we sort them in ascending order and denote the $i^{th}$ smallest number by $R_i$,(i.e., we have $R_1 < R_2 < \ldots < R_m$). Let $\BetaD(\alpha,\beta)$ denote the Beta distribution\footnote{The Beta distribution is a family of continuous probability distributions defined on the interval $0 \leq x \leq 1$ with shape parameters $\alpha$ and $\beta$, and with probability density function $f(x; \alpha,\beta) = \frac{x^{\alpha-1}(1-x)^{\beta-1}}{B(\alpha,\beta)}$, where the constant  $B(\alpha,\beta) = \frac{\Gamma(\alpha)\Gamma(\beta)}{\Gamma(\alpha+\beta)}$ and $\Gamma(z) = \int_{0}^{\infty} t^{z-1}e^{-t} \mathbf{d}t$ is the Gamma function.}. Then, for an arbitrary $R_{m+1}$ drawn from the same
distribution $\mathcal{D}$, the following holds:
\begin{equation}
\label{eq:beta}
\Pr\left[ R_{m+1} < R_\ell \right] \sim \BetaD(\ell , m+1-\ell), \quad 1\leq \ell \leq m.
\end{equation}
\end{lemma}
The original $m$ i.i.d. data-points are called a {\em calibration set}. The
above lemma says that \navidd{the probability for } a {\em previously unseen} data-point $R_{m+1}$ drawn from the same distribution $\mathcal{D}$ being less than the $\ell^{th}$ smallest number in the calibration set is itself a random variable that has a specific Beta distribution. We next show how we can exploit this lemma to obtain probabilistic correctness guarantees for our trained controllers.

We assume that there is some user-specified distribution over the set of initial states in $\init$, and that we can sample $m$ initial states
$\statee_{0,1},\ldots,\statee_{0,m}$ from this distribution. For a sampled
initial state $\statee_{0,i}, i \in  1,\cdots,m$, we can obtain the corresponding negative robust satisfaction value, and set: $R_i = -\rob(\varphi, \traj[\statee_{0_i}\ ;\param],0), i\in 1,\cdots,m.$

From Lemma~\ref{lem:vovk}, we know that for a previously unseen initial state $\statee_{0,m+1}$, the corresponding (negative \navid{value of}) robustness $R_{m+1}$ satisfies the relation in \eqref{eq:beta}. Now, almost all sampled trajectories \navidd{generated by} a trained controller are expected to have positive robustness value, so we expect the quantities $R_1,\ldots,R_m$ to be all negative. In the pessimistic case, we expect at least the first $\ell$ of these quantities to be negative. If so, the guarantee in Eq.~\eqref{eq:beta} essentially quantifies the probability of the robustness of a trajectory for a previously unseen initial state to be positive. Note that:
\begin{eqnarray}
& (R_{m+1} < R_\ell) \wedge (R_\ell < 0) \implies (R_{m+1} < 0) \implies 
(\traj[\statee_{0,m+1}\ ;\param] \models \varphi) & \\
& \therefore \Pr(\traj[\statee_{0,m+1}\ ;\param] \models \varphi) \ge \Pr(R_{m+1} < R_\ell) \sim \BetaD(\ell,m+1-\ell) \label{eq:bigger}
\end{eqnarray}

\navidg{In addition, from \cite{david2004order} we know that the mean and variance of the Beta distribution are given as follows}:
\begin{equation}\label{eq:beta_features}
\mathbb{E}\big[\Pr[R_{m+1}<R_\ell]\big] = \frac{\ell}{m+1} \quad 
\mathbf{Var}\big[\Pr[R_{m+1}<\navid{R_\ell}]\big] = \frac{\ell(m+1-\ell)}{(m+1)^2(m+2)}.
\end{equation}
\navidd{As the Beta distribution has small variance and is noticeably sharp}, the desired coverage level for a probabilistic guarantee can be obtained \navidd{in the vicinity of its mean value. From the closed form formula in \eqref{eq:beta_features}, we observe that in case} we wish to have a coverage level \navidd{close to} $(1-10^{-4})$ or $99.99\%$, then we can set $\ell = \lceil(m+1)(1-10^{-4})\rceil$. Here we also set $m$ to $10^5$, giving the value of $\ell = 99991$.  Let's denote $\Pr[R_{m+1} < R_\ell]$ as $\delta$. Since $\delta$ is a random variable sampled from $\mathbf{Beta}(\ell , m +1-\ell)$ where ($\ell = 99991, m= 10^5$)\footnote{We can compute for its mean and variance via \eqref{eq:beta_features} as $\mu = \mathbb{E}[\delta] = 0.9999$ and $\mathbf{var}[\delta] = 9.9987 \times 10^{-10}$.}, we can utilize the cumulative density function of Beta distribution (i.e, regularized incomplete Beta function) and for a given $\delta_1\in (0,1)$ propose the following guarantee, 
$$
\Pr[ \delta \geq \delta_1] = 1 - I_{\delta_1}(\ell , m+1-\ell),\  
$$
where  $I_x(\ .\ ,\ .\ )$ is the regularized incomplete Beta function at point $x$.

Here $\delta_1$ is the desired confidence level that we consider for the probabilistic guarantee. However, if we set $\delta_1=0.9999$ then $ \Pr[ \delta \geq 0.9999] = 0.54$ which indicates that the confidence in the $99.99$\% guarantee is low. If we instead set $\delta = 0.9998$, this results in 
$\Pr[ \delta \geq 0.9998] = 0.995$, which indicates a much higher level of confidence. Finally, based on \eqref{eq:bigger}, we can consider the provided guarantee also for the trajectories and conclude,

\begin{equation}\label{eq:finalguarantee}
\Pr[\ \ \ \Pr[\traj[\statee_0\ ;\param] \models \varphi]\ \ \  \geq 99.98\%\ \ \ ]\ \ \  \geq 99.5\%
\end{equation} 
To summarize, in each of our case studies, we sample $m = 10^5$ i.i.d.
trajectories, compute their sorted negative robustness values $R_1,\ldots,R_m$,
and check that $R_\ell$ for $\ell = 99991$ is indeed negative. This gives us the
probabilistic guarantee provided in \eqref{eq:finalguarantee} that from unseen
initial conditions the system will not violate the DT-STL specification.


\section{Related Work and Conclusion}
\label{sec:conc}
\mypara{Related Work}
In the broad area of formal methods, robotics, and cyber-physical systems, there
has been substantial research in synthesizing controllers from temporal logic
specifications. This research involves different considerations. First, the
plant dynamics may be specified as either a differential/difference
equation-based model \citep{gilpin2020smooth,pant_iccps,raman2014model,farahani2015robust,lindemann2018control,raman2015reactive,guo2018probabilistic},
or as a Markov decision process \citep{sadigh2016safe,haesaert2018temporal,kalagarla2020synthesis} that
models stochastic uncertainty, or may not be explicitly provided (but is
implicitly available through a simulator that samples model behaviors). The
second consideration is the expressivity of the specification language, i.e., if
the specifications are directly on the real-valued system behaviors or on
Boolean-valued propositions over system states, and if the behaviors are over a
discrete set of time-steps or over dense time. Specification languages such as
LTL (Linear Temporal Logic) \cite{pnueli1977temporal}, Metric Temporal Logic
(MTL) \cite{koymans1990specifying} and Metric Interval Temporal Logic (MITL)
\cite{alur1991techniques} are over Boolean signals, while Signal Temporal Logic
(STL) \cite{maler2004monitoring} and its discrete-time variant DT-STL considered
in this paper are over real-valued behaviors. MTL, MITL and STL are typically
defined over dense time signals while LTL and DT-STL are over discrete
time-steps. The third consideration is the kind of controller being synthesized.
Given the plant dynamics, some techniques find the entire sequence of control
actions from an initial state to generate a desired optimal trajectory (open
loop control) \cite{yaghoubi2019gray,raman2014model,pant2017smooth,lindemann2018control}, while
some focus on obtaining a feedback controller that guarantees satisfaction of
temporal logic objectives in the presence of uncertainty
 \cite{YaghoubiF2019tecs} (in the initial states or during system
execution).  We now describe some important sub-groups of techniques in this
space that may span the categories outlined above.

\myipara{Reactive Synthesis}  A reactive synthesis approach models the system
interaction with its environment as a turn-based game played by the system and
the environment over a directed graph \cite{bloem2018graph}. The main idea is to
convert temporal logic specifications (such as LTL) into winning conditions and
identify system policies that deterministically guarantee satisfaction of the
given specification \cite{kress2009temporal}. As reactive synthesis is a
computationally challenging problem, there are many sub-classes and heuristics
that have been explored for efficiency; for instance, in
\cite{wongpiromsarn2012receding} a receding horizon framework is used; in
\cite{tuumova2010symbolic}, the authors focus on piece-wise affine
nondeterministic systems, while \cite{raman2015reactive} investigates reactive
synthesis for STL. 



\myipara{Reinforcement and Deep Reinforcement Learning} Reinforcement learning
(RL) algorithms learn control policies that maximize cumulative rewards over
long-term horizons. Recently, RL temporal has been used to infer reward
functions that can guarantee satisfaction of an LTL specification
\cite{hasanbeig2018logically,sadigh2014learning,bozkurt2020control,fu2014probably}.
The work in
\cite{venkataraman2020tractable,kapoor2020model,hamilton2022training,puranic2022learning,balakrishnan2019structured}
generate reward functions from STL specifications. While the ultimate objective
of these methods is similar to our problem setting, we adopt a model-based
approach to control synthesis where we \navid{assume access to a differentiable model of the system and
use gradient ascent to train the controller} in contrast to RL algorithms that
may rely on adequate exploration of the state space to obtain near-optimal
policies (that may guarantee satisfaction of specifications).


\myipara{MPC and MILP} A clever encoding of LTL as mixed integer linear (MIL)
constraints was presented in \cite{wolff2014optimization} for the purpose of
reactive synthesis. This idea was then extended in \cite{raman2014model} to show
that model predictive control of linear/piecewise affine systems w.r.t. STL
objectives (with linear predicates) can be solved using mixed integer linear
programming (MILP) solvers. MILP is an NP-hard problem, and various optimization
improvements to the orignal problem
\cite{kurtz2022mixed,gilpin2020smooth,takayama2023signal,pant2017smooth} and extensions to
stochastic systems  \cite{farahani2015robust,sadraddini2015robust} have been
proposed. In contrast to a model-predictive controller, we obtain a NN feedback
controller that does not require online optimization required in MPC.

%


\myipara{Barrier Function-based Approaches} A control barrier function (CBF) can be thought of as a safety envelope for a controlled dynamical system. As long as the CBF satisfies validity conditions (typically over its Lie derivative), the CBF guarantees the existence of a control policy that keeps the overall system safe \cite{xu2015robustness}. CBFs can be used to enforce safety constraints and also to enforce temporal specifications such as STL \cite{lindemann2018control,ames2016control,ames2014control,deshmukh2019learning}. The design of barrier functions is generally a hard problem, though recent research studies compute for the CBFs through 
learning \cite{venkataraman2020tractable,robey2021learning}, and using quantitative semantics of STL \cite{hashemi2023risk}.


\myipara{Gradient-based Optimization methods} This class of methods investigates
learning neural network controllers by computing the gradient of the robustness
function of STL through back-propagation  STL. For instance, training feedback
neural network controllers is studied in \cite{HashemiEtAl2023acc,
YaghoubiF2019tecs, yaghoubi2019gray, liu2021recurrent, hashimoto2022stl2vec} and
for open-loop controllers is investigated in \cite{leung2023backpropagation}.
The main contributions in this paper over previous work is to scale gradient
descent to long time-horizons using the novel idea of dropout, and a more
efficient (and smooth) computation graph for STL quantitative semantics.

\myipara{Prior work on NN controllers for STL} The overall approach of this
paper is the closest to the work in
\citep{YaghoubiF2019tecs,leung2019backpropagation,LeungAP2021wafr,hashemi2023neurosymbolic,HashemiEtAl2023acc},
where STL robustness is used in conjunction with back-propagation to train
controllers. The work in this paper makes significant strides in extending
previous approaches to handle very long horizon temporal tasks, crucially
enabled by a novel sampling-based gradient approximation. Due to the structure
of our NN-controlled system, we can seamlessly handle time-varying dynamics and
complex temporal dependencies. We also note that while some previous approaches
focus on obtaining open-loop control policies, we focus on synthesizing
closed-loop, feedback NN-controllers which can be robust to minor perturbations
in the system dynamics. In addition, we cover a general DT-STL formula for
synthesis, and we utilize $\lbfortl$ \cite{hashemi2024scaling} for backward
computation that has shown significant improvement for efficiency of training
over complex DT-STL formulas.   

\myipara{Limitations} Some of the key limitations of our approach include: (1)
we
do not address infinite time horizon specifications. (2) We only consider a
discrete-time variant of STL. (3) Our approach would fail if the chosen neural
network architecture for the controller has too few parameters (making it
difficult to control highly nonlinear environment dynamics) or if it has too
many parameters (making it a difficult optimization problem). (4) We assume
full system observability and do not consider stochastic dynamics.

\mypara{Conclusion} Using neural network feedback controllers for control
synthesis offers robustness against noise and uncertainties, making them
preferable over open-loop controllers. However, training these controllers can
be challenging due to issues like vanishing or exploding gradients, especially
in long time horizons or high-dimensional systems. To address this challenge, we
introduced a gradient sampling technique inspired by dropout
\cite{srivastava2014dropout} and stochastic depth \cite{huang2016deep}.
Additionally, we proposed incorporating critical predicates into this technique
to enhance training efficiency, and we tested our approach on various
challenging control synthesis problems.

\bibliography{main}


\end{document}